\documentclass[aps,prl,twocolumn,superscriptaddress]{revtex4-2} 
\usepackage{graphicx}  
\usepackage{epstopdf}
\usepackage{dcolumn}   
\usepackage{bm}        
\usepackage{geometry}
\usepackage{amsmath}   
\usepackage{amssymb}   
\usepackage{mathtools}
\usepackage{maybemath}
\usepackage{xcolor}
\usepackage{enumitem}
\usepackage{graphicx}  
\usepackage{nicefrac}
\usepackage{comment}
\usepackage{physics}
\newcommand{\muB}[0]{\mu_\textrm{B}}
\newcommand{\stat}[1]{\langle #1 \rangle}%
\newcommand{\statop}[1]{\langle \hat{#1} \rangle}%
\newcommand{\rarr}[0]{\rightarrow}
\newcommand{\const}[0]{\textrm{const}}
\newcommand{\eps}[0]{\varepsilon}

\newcommand{\up}[0]{\uparrow}
\newcommand{\dn}[0]{\downarrow}
\newcommand{\dw}[0]{\downarrow}
\newcommand{\ho}[0]{\mathrm{ho}}
\newcommand{\lu}[0]{\mathrm{lu}}
\newcommand{\lp}[0]{\left}
\newcommand{\rp}[0]{\right}
\newcommand{\s}[0]{\sigma}

\newcommand{\de}[0]{\partial}
\newcommand{\KS}[0]{\mathrm{KS}}

\newcommand{\leqs}[0]{\leqslant}
\newcommand{\geqs}[0]{\geqslant}

\newcommand{\half}[0]{\frac{1}{2}}
\newcommand{\thalf}[0]{\tfrac{1}{2}}
\newcommand{\rr}[0]{\mathbf{r}}

\newcommand{\tot}{\textrm{tot}}
\newcommand{\m}[0]{\textrm{min}}
\newcommand{\ens}{\textrm{ens}}

\newcommand\identity{1\kern-0.25em\text{l}}
\newcommand{\defeq}{\vcentcolon=}

\begin{document}

\title{Ensemble ground-state of a many-electron system with fractional electron number and spin: piecewise-linearity and flat plane condition generalized}

\author{Yuli Goshen}
\affiliation{Fritz Haber Research Center for Molecular Dynamics and Institute of Chemistry, The Hebrew University of Jerusalem, 9091401 Jerusalem, Israel}

\author{Eli Kraisler}
\email{eli.kraisler@mail.huji.ac.il}
\affiliation{Fritz Haber Research Center for Molecular Dynamics and Institute of Chemistry, The Hebrew University of Jerusalem, 9091401 Jerusalem, Israel}

\date{\today}

\begin{abstract}

Description of many-electron systems with a fractional electron number $N_\tot$ and fractional spin $M_\tot$ is of great importance in physical chemistry, solid state physics and materials science.
In this Letter, we provide an exact description of the zero-temperature ensemble ground state of a general, finite, many-electron system, and characterize the dependence of the energy and the spin-densities on both $N_\tot$ and $M_\tot$, when the total spin is at its equilibrium value. 
We generalize the piecewise-linearity principle and the flat-plane condition and determine which pure states contribute to the ground-state ensemble. We find a new derivative discontinuity, which manifests for spin variation at constant $N_\tot$, as a jump in the Kohn-Sham potential. We identify a previously unknown degeneracy of the ground state, such that the total energy and density are unique, but the spin-densities are not. Our findings serve as a basis for development of advanced approximations in density functional theory and other many-electron methods.

\end{abstract}
\maketitle{}

Modelling of materials and chemical processes from first principles crucially depends on efficient, accurate, yet approximate methods (see, e.g., ~\cite{Szabo,PY,DG,Primer,FetterWalecka,Reining_MB}).
The quality of a given approximation can be assessed not only versus the experiment, but also by comparison to known exact properties of many-electron systems; in the context of density functional theory (DFT), see, e.g.,~\cite{Perdew05,Perdew09_perplexed,KaplanLevyPerdew23} One such exact property is the piecewise-linear behavior of the energy versus electron number~\cite{PPLB82}: when varying the number of electrons in the system, $N_\tot$, in a continuous manner, allowing both integer and fractional values, the energy $E(N_\tot)$ is linear between any two integer values of $N_\tot$. As $N_\tot$ surpasses an integer, the slope of $E(N_\tot)$ may abruptly change. 
\emph{Piecewise linearity} exists also for the electron density $n(\rr)$ and any property that is an expectation value of an operator. 

Spin, which is a fundamental property of an electron, 
has to be taken into account in many-electron systems, both when magnetic fields are present, but also in their absence\cite{Heinrich04,CapVigUll10,CasanovaKrylov20}. 
Occurrence of fractional $z$-projection of the total spin, $M_\tot$, and the dependence of the energy on $M_\tot$, has been extensively studied in the literature\cite{Chan99,CohenMoriSYang08, MoriS09, GalGeerlings10JCP, GalGeerlings10PRA, cohen12, Saavedra12, GouldDobson13, XDYang16}.
In Ref.~\citenum{CohenMoriSYang08} it was shown that the exact total energy must be constant for all $M_\tot \in [-S_\m, S_\m]$ (where $S_\m$ is the \emph{equilibrium} value of the \emph{total} spin, $S$, i.e., that value of $S$ for which the system has the lowest energy). 
In Ref.~\citenum{MoriS09} the  \emph{flat-plane condition} has been developed, unifying piecewise-linearity~\cite{PPLB82} and the constancy condition~\cite{CohenMoriSYang08}, which are both completely general principles, and apply to any many-electron approach. Focusing on the H atom, Ref.~\citenum{MoriS09} described its exact total energy, while continuously varying $N_\tot$ from 0 to 2 and $M_\tot$ accordingly, while $M_\tot \in [-\tfrac{1}{2}, \tfrac{1}{2}]$.%
~\footnote{Hartree atomic units are used throughout}  %
The exact energy graph of H in the $N_\up - N_\dw$ plane (where $N_\up = \tfrac{1}{2} N_\tot + M_\tot$ and $N_\dw = \tfrac{1}{2} N_\tot - M_\tot$)
is comprised of two triangles whose vertices lie at the points of integer $N_\up$ and $N_\dw$. 
Approximate exchange-correlation (xc) functionals may strongly violate this condition, leading to delocalization error and static correlation error. 

These developments were followed by numerous studies (see, e.g., \cite{Cohen08,BajajKulik17,WodynskiArbuzKaupp21,Prokopiu22,BajajKulik22,BurgessLinscottORegan23} and references therein) which focused on: (a) the constancy condition~\cite{CohenMoriSYang08}, examined by changing $M_\tot$ between $-S_\m$ and $+S_\m$ while keeping $N_\tot$ constant and integer; or (b) varying both $N_\tot$ and $M_\tot$ for systems with $S_\m = \tfrac{1}{2}$. 
In addition, Refs.~\citenum{Chan99, GalGeerlings10JCP, Saavedra12} and~\citenum{XDYang16} analyzed the flat-plane behavior of the energy for general $N_\up$ and $N_\dn$, suggesting formulae describing the energy profile.
They assumed that the total energy (and analogously other properties) for any fractional $N_\tot$ and $M_\tot$ is determined by the energies of three out of the four states with nearby integer numbers of $\up$ and $\dw$ electrons. The overall energy graph is therefore a collection of triangles. Moreover, Refs.~\citenum{GalAyersProftGeerlings09} and~\citenum{GalGeerlings10PRA} raised the possibility of a discontinuity in $E$ versus $M_\tot$, which stems from the nonuniqueness of the external magnetic field in spin-DFT.

This work goes beyond previous results, and rigorously describes the exact ground state of a general, finite, many-electron system with an arbitrary fractional electron number, $N_\tot$, and a simultaneously fractional $z$-projection of the spin $M_\tot$, as long as the total spin, $S$, is at its equilibrium value, $S = S_\m$; $S_\m$ itself can be of any value. Such a ground state has to be described by an ensemble, attained by minimizing the total energy, while constraining $N_\tot$ and $M_\tot$. 

Our findings generalize the piecewise-linearity principle~\cite{PPLB82} and the flat-plane condition~\cite{MoriS09}. We find which pure ground states contribute to the ensemble, and which do not. The contributing pure states are not necessarily among the four neighboring states to the point $(N_\tot,M_\tot)$, contrary to previous findings. %
Notably, the energy graph consists not only of triangles, but also of trapezoids. 
Furthermore, we identify a new type of a derivative discontinuity, which manifests in the case of spin variation, as a jump in the Kohn-Sham (KS) potential and compare it to the discontinuities defined in Ref.~\citenum{CapVigUll10}. 
Surprisingly, we reveal a degeneracy in the ensemble ground state, where the energy and the total density are determined uniquely, but the spin-densities are not. Our findings provide new exact properties of many-electron systems, useful for development of advanced approximations in DFT and in other many-electron methods.

We start our derivation by considering  \emph{an ensemble} ground state $\hat{\Lambda}$ of a system with a possibly fractional $N_\tot$ and fractional $M_\tot$, where no magnetic fields are present.
The (non-relativistic) Hamiltonian is given by $\hat{H}=\hat{T}+\hat{V}+\hat{W}$, where $\hat{T}$ is the kinetic energy operator, $\hat{V}$ is a multiplicative external potential operator and $\hat{W}$ is the electron-electron repulsion operator.
Subsequently, $\hat{\Lambda}$ minimizes the expectation value of the energy, $\Tr{\hat{\Lambda}\hat{H}}$, under the constraints
\begin{equation}
\label{constraint_N}
\stat{\hat{N}} \defeq \Tr{\hat{\Lambda}\hat{N}}=N_\tot,
\end{equation}
\begin{equation}
\label{constraint_S}
\stat{\hat{S}_{z}} \defeq \Tr{\hat{\Lambda}\hat{S}_z}=M_\tot
\end{equation}
and
\begin{equation}
\label{constraint_prob}
      \stat{\hat{\identity}}  \defeq \Tr\hat{\Lambda}=1.
\end{equation}
Here $\hat{N}$ is the electron number operator and $\hat{S}_z$ is the operator for the $z$-projection of the total spin. 

Consider now $\ket{\Psi_{N,S,M}}$,
the lowest-energy eigenstate of the operators $\hat{N},\hat{S}^2,\hat{S}_z$, with eigenvalues $N$, $S(S+1)$ and $M$, respectively. $N$ is a non-negative integer, $M$ and $S$ are integers (for even $N$) or half-integers (for odd $N$), and $-S\leqs M \leqs S$.
For a given value of $N$, there exists one such value of $S$ that minimizes the energy; we denote it $S_\m$. Subsequently, there exists a multiplet of $(2S_\m+1)$ states, all with the same energy (in absence of a magnetic field), and with $M = -S_\m, \ldots, S_\m$. We assume that for all other values of $S$ (each one with its own multiplet of states) the energy is strictly higher. We further assume, for simplicity, that for specified $N$, $S$ and $M$, the pure ground state is not degenerate any further. 

\begin{figure}
    \centering
    \includegraphics[width=1\linewidth]{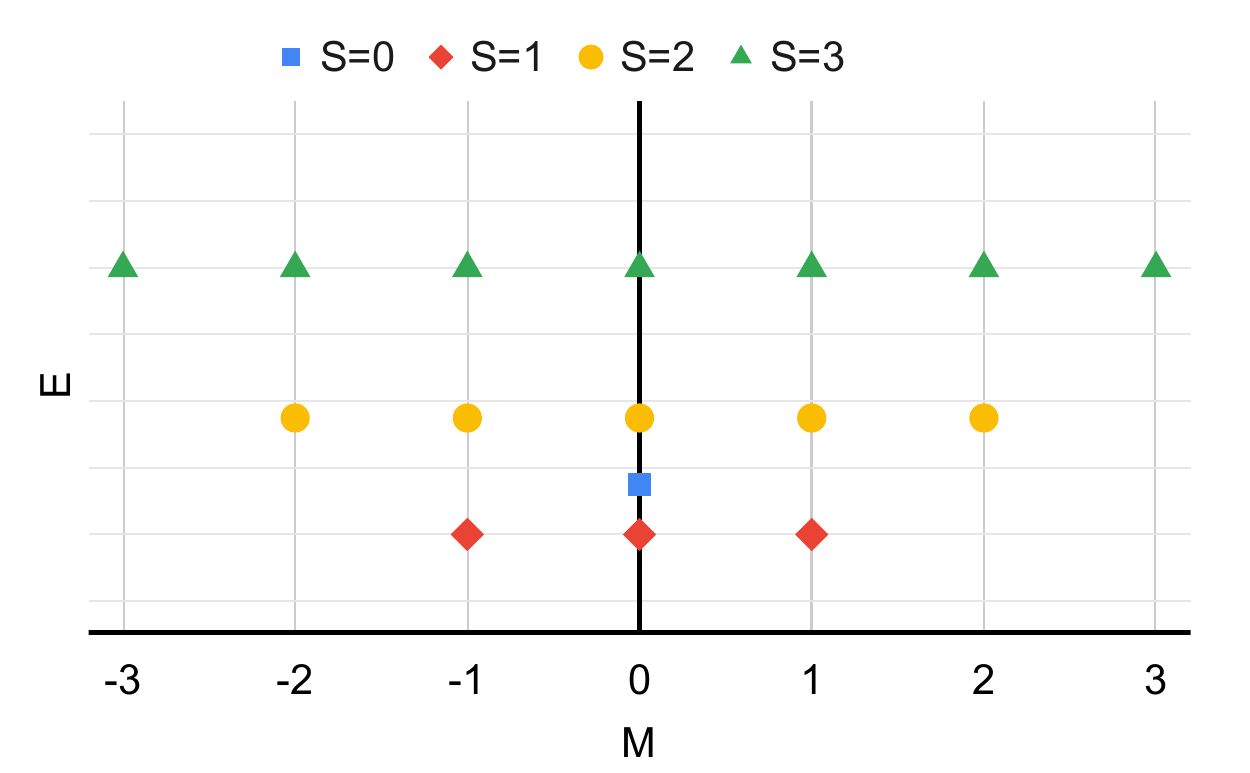}
    \caption{ 
Illustration of the energy $\expval{\hat{H}}{\Psi_{N,M,S}}$ versus $M$, at constant $N$, for various values of $S$ (see legend). In this case, $S_\m = 1$.}
    \label{fig:spin_energies}
\end{figure}

In the following we denote $\ket{\Psi_{N,M}}$ as the lowest-energy eigenstate of $\hat{N}$ and $\hat{S}_z$, with eigenvalues $N$ and $M$, respectively. For $M \in [-S_\m, S_\m]$, we have, of course, $\ket{\Psi_{N,M}}=\ket{\Psi_{N,S_\m(N),M}}$; varying $M$ does not affect the energy $E(N,M)=\expval{\hat{H}}{\Psi_{N,M}}$. For other values of $M$, the ground state energy is higher. The energy $\expval{\hat{H}}{\Psi_{N,S,M}}$ as a function of $S$ and $M$ is as illustrated in Fig.~\ref{fig:spin_energies}. 
Moreover, as long as $M\in[-S_\m(N),S_\m(N)]$, the density $n_N(\rr)=\expval{\hat{n}(\rr)}{\Psi_{N,M}}$ does not depend on $M$ (see Supporting Information (SI) for details).

For general, possibly fractional, values of $N_\tot$ and $M_\tot$,
the ground state $\hat{\Lambda}$ is expressed as $\hat{\Lambda}=\sum_{N}\sum_{M}\lambda_{N,M}\ketbra{\Psi_{N,M}}$, where $\lambda_{N,M} \in [0,1]$.%
~\footnote{We use here the fact that $\hat{N}$, $\hat{S}^2$, $\hat{S}_z$,  and $\hat{H}$ are commuting operators and therefore their eigenstates, $\ket{\Psi_{N,S,M,E}}$, form a complete basis. Among these, only $\ket{\Psi_{N,M}}$ are necessary to describe the ground state $\hat{\Lambda}$, as they have the lowest energy for a given $N$ and $M$. We note in passing that more general ensembles, which include also off-diagonal terms of the form  $\ketbra{\Psi_{N_a,M_a}}{\Psi_{N_b,M_b}}$, introduce additional degeneracy to the ground state, but the energy and the spin-densities $n^\sigma_\ens(\rr)$ are unaffected by this extension.}
Here and below the sum over $N$ includes all integers for which the $N$-electron system is bound, and the sum over $M$ includes all possible (integer or half-integer) values of $M$, for a given value of $N$: $M = -\tfrac{N}{2}, -\tfrac{N}{2}+1, \ldots, \tfrac{N}{2}-1, \tfrac{N}{2}$.
In the following, we denote $N_\tot=N_0+\alpha$, where $N_0$ is the integer part and $\alpha \in [0,1)$ is the fractional part of $N_\tot$.
The equilibrium spin values for $N_0$ and $N_0+1$ electrons are, respectively, $S_0 \defeq S_\m(N_0)$ and $S_1 \defeq S_\m(N_0+1)$.

We now consider the important case of 
\begin{equation} \label{eq:spin_condition}
|M_{\tot}| \leqs (1-\alpha)S_0+\alpha S_1 
\end{equation} 
and prove that \emph{any} ensemble state, which consists of only 
$\ket{\Psi_{N_0,-S_0}},\ldots,\ket{\Psi_{N_0,S_0}}$ and  $\ket{\Psi_{N_0+1,-S_1}}, \ldots,$ $\ket{\Psi_{N_0+1,S_1}}$,
and satisfies Constraints \eqref{constraint_N}--\eqref{constraint_prob}, is \emph{a ground state} of the system. 

By explicitly writing the aforementioned ensemble state as 
\begin{align} \label{gamma}    
& \hat{\Gamma}=\sum_{N=N_0}^{N_0+1}\sum_{M = - S_{\m}(N)} ^{S_\m(N)} \gamma_{N,M}\ketbra{\Psi_{N,M}}
\end{align}
and making use of Eqs.~\eqref{constraint_N} and \eqref{constraint_prob} (for $\hat{\Gamma}$), we see that
$\sum_{M = -S_0}^{S_0} \gamma_{N_0, M}=1-\alpha$ and $\sum_{M=-S_1}^{S_1} \gamma_{N_0+1, M}=\alpha$. 
It then follows that the energy of this ensemble state is $\Tr{\hat{\Gamma}\hat{H}}=(1-\alpha)E(N_0)+\alpha E(N_0+1)$, where $E(N) \defeq \min_M E(N,M) = E(N,S_{\m}(N))$ is the 
ground state energy for a given $N$. Notably, there is no dependence on the $z$-projection of the spin;  Constraint~\eqref{constraint_S} was not used. Now we show that $\Tr{\hat{\Gamma}\hat{H}}$ is indeed the ground state energy. 

First, from the piecewise-linearity principle of Ref.~\citenum{PPLB82}, it follows that $ \min_{\hat{\Xi} \mapsto N_\tot} \Tr{\hat{\Xi}\hat{H}}=(1-\alpha)E(N_0)+\alpha E(N_0+1)$, namely that the ground state energy for any system with $N_\tot$ electrons, represented by a state $\hat{\Xi}$, is linear in $N$, between $N_0$ and $N_0+1$.
Second, note that $E_\ens(N_\tot,M_\tot) 
\defeq \min_{\hat{\Xi}\mapsto N_\tot, M_\tot} \Tr{\hat{\Xi}\hat{H}} \geqs  \min_{\hat{\Xi}\mapsto N_\tot}\Tr{\hat{\Xi}\hat{H}} $, merely stating the ensemble ground-state energy for given  $N_\tot$ and $M_\tot$ is greater or equal to the ensemble ground-state energy given only $N_\tot$, because adding an additional constraint in a minimization yields a result larger or equal the original one.
Third, since $\hat{\Gamma}$ satisfies Eqs.~\eqref{constraint_N}--\eqref{constraint_prob}, meaning that $\hat{\Gamma}$ is an ensemble that yields $N_\tot$ and $M_\tot$,
$\Tr{\hat{\Gamma}\hat{H}} \geqs \min_{\hat{\Xi}\mapsto N_\tot, M_\tot} \Tr{\hat{\Xi}\hat{H}}$. 
Combining all the above, we find that 
\begin{align}
& \min_{\hat{\Xi}\mapsto N_\tot, M_\tot} \Tr{\hat{\Xi}\hat{H}} \leqs \Tr{\hat{\Gamma}\hat{H}} =  (1-\alpha) E(N_0) + \nonumber \\ 
& + \alpha E(N_0+1) =  \min_{\hat{\Xi} \mapsto N_\tot}\Tr{\hat{\Xi}\hat{H}} \leqs \min_{\hat{\Xi} \mapsto N_\tot, M_\tot} \Tr{\hat{\Xi}\hat{H}}. 
\end{align} 
This inequality holds only if all its terms equal each other, meaning that $\Tr{\hat{\Gamma}\hat{H}}=\min_{\hat{\Xi} \mapsto N_\tot, M_\tot} \Tr{\hat{\Xi}\hat{H}}$, i.e., $\hat{\Gamma}$ is indeed \emph{a} ground state.  

Next, to show that \emph{every} ground-state is of the form of $\hat{\Gamma}$ (Eq.\eqref{gamma}), still within Region~\eqref{eq:spin_condition}, we assume, by way of contradiction, that this is false for a ground state $\hat{\Lambda}$, which is explicitly written as $\hat{\Lambda}=\sum_N \sum_{M} \lambda_{N,M}\ketbra{\Psi_{N,M}}$. Next, we define an auxiliary quantity,
\begin{equation}\label{symmetrize}
 \hat{\Lambda}^\prime \defeq \sum_N \sum_{M} \frac{\lambda_{N,M}}{2}\left(\ketbra{\Psi_{N,M}}+\ketbra{\Psi_{N,-M}}\right),
\end{equation}
to deduce properties of $\hat{\Lambda}$.
The sum on $M$ in Eq.~\eqref{symmetrize} runs over both positive and negative values, as before.
$\hat{\Lambda}'$ satisfies Eqs.~\eqref{constraint_N} and \eqref{constraint_prob}, $\Tr{\hat{\Lambda}^\prime \hat{S}_z}=0$, and $\Tr{\hat{\Lambda}^\prime \hat{H}}=\Tr{\hat{\Lambda} \hat{H}}=(1-\alpha)E(N_0)+\alpha E(N_0+1)$. 
Therefore, $\hat{\Lambda}^\prime$ is itself a ground state, of $N_{\tot}$ electrons and spin $0$. 

To find the coefficients $\lambda_{N,M}$ that bring the energy of $\hat{\Lambda}^\prime$ to minimum, while retaining its expectation values of $\hat{\identity},\hat{N},\hat{S}_z$, we notice that, \emph{for each value of $N$} (in the external sum of Eq.~\eqref{symmetrize}), the energy is minimized by setting $\lambda_{N,M}=0$ for all $M>S_{\m}(N)$ and $M<-S_{\m}(N)$.
Then, $\hat{\Lambda}^\prime=\sum_{N}  \sum_{M=-S_{\m}(N)} ^{S_\m(N)}\frac{\lambda_{N,M}}{2} (\ketbra{\Psi_{N,M}}+\ketbra{\Psi_{N,-M}})$ and $\Tr{\hat{\Lambda}^\prime \hat{H}}=\sum_N  l_N E(N)$ where $l_N=\sum_{M=-S_\m(N)}^{S_\m(N)}\lambda_{N,M}$. 
This expression for the ensemble energy is familiar from Ref.~\citenum{PPLB82}: Using the convexity conjecture of the energy\cite{DG,Lieb,cohen12}, we state that this energy is minimized by
\begin{equation} \label{eq:l_N_contradiction}
l_N=\begin{cases}
			1-\alpha
            & \text{if $N=N_0$}\\
            \alpha &
            \text{if $N=N_0+1$}\\ 
            0 & \text{otherwise}
		 \end{cases},    
\end{equation}
still under the aforementioned constraints for $\hat{\Lambda}^\prime$. This means that the only nonzero coefficients of $\hat{\Lambda}$ are $\lambda_{N_0,-S_0},\ldots,\lambda_{N_0,S_0}$ and $\lambda_{N_0+1,-S_1},\ldots,\lambda_{N_0+1,S_1}$, in contradiction to the original assumption. 

This concludes our proof: the ground state of a system with $N_\tot$ electrons with a total spin $M_\tot$ is described \emph{only} by the ensemble ground state $\hat{\Gamma}$ of Eq.~\eqref{gamma}. It consists of  $2(S_0+S_1+1)$ pure states, at most. The coefficients $\gamma_{N,M} \in [0,1]$ have to be determined from Constraints \eqref{constraint_N}--\eqref{constraint_prob}, which can be expressed as:

\begin{align} \label{eq:gamma_coeffs} 
& \sum_{M=-S_0} ^{S_0} \gamma_{N_0,M} = 1-\alpha; 
\sum_{M = -S_1}^{S_1} \gamma_{N_0+1,M} = \alpha; \nonumber \\
& \sum_{M=-S_0} ^{S_0} M \gamma_{N_0,M} + \sum_{M=-S_1} ^{S_1} M \gamma_{N_0+1,M} = M_\tot.
\end{align}

Several important consequences follow from the fact that $\hat{\Gamma}$ of Eq.~\eqref{gamma} is a ground state. 

\begin{figure}
    \centering
\includegraphics[width=0.45\textwidth]{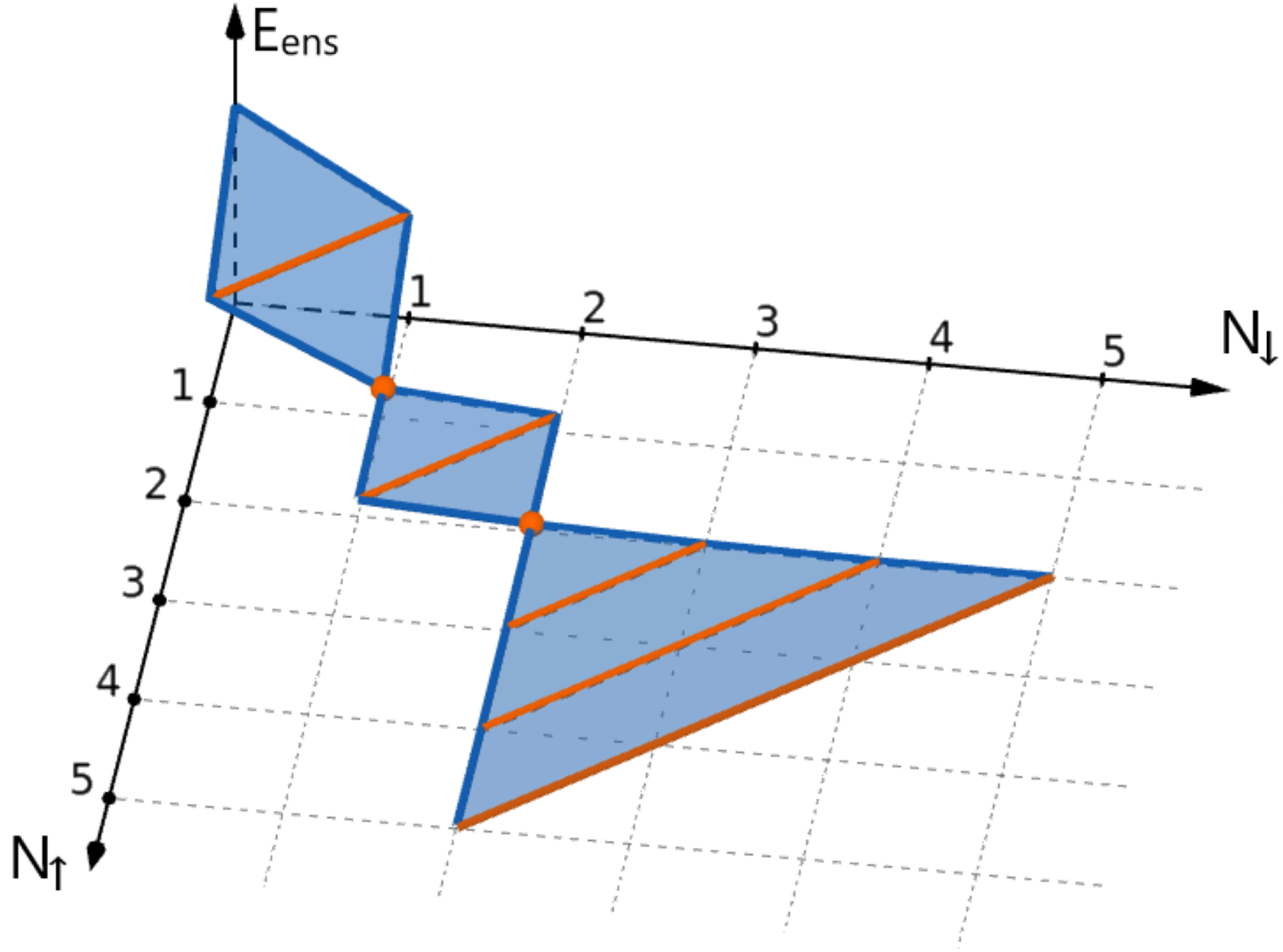}
    \caption{
The ground-state energy $E_\ens(N_\up,N_\dn)$ for the C atom, within the region defined in Eq.~\eqref{eq:spin_condition}, and for $0 \leqs N_\tot \leqs 7$. 
    The slope is zero along the orange line segments, which correspond to constant and integer $N_\tot$, and changes discontinuously in directions perpendicular to these lines. }
    \label{fig:Nup_Ndw}
\end{figure}

\noindent  \emph{1.\ Total energy.}  Within Region~\eqref{eq:spin_condition}, the total energy 
\begin{equation}
E_\ens(N_\tot,M_\tot) = \Tr{\hat{\Gamma}\hat{H}} = (1-\alpha) E(N_0) + \alpha E(N_0+1)   
\end{equation}
is piecewise-linear with respect to $\alpha$ and does not depend on the $z$-projection of the spin, $M_\tot$. The graph of $E_\ens(N_\up,N_\dn)$ forms a series of triangles and trapezoids, as presented in Fig.~\ref{fig:Nup_Ndw}. 
For the H atom, with the number of electrons in each spin channel varying from 0 to 1, this result is precisely the flat-plane condition as described in Ref.~\citenum{MoriS09}. However, for larger $N_\tot$ and particularly for
$S_{\min} > \tfrac{1}{2}$, e.g., the C atom of 
Fig.~\ref{fig:Nup_Ndw}, the behavior of $E_\ens$ is richer.
Furthermore, since $E_\ens$ does not depend on the spin, the slope in Fig.~\ref{fig:Nup_Ndw} along lines of constant $N_\tot$ is $(\de E_\ens / \de M_\tot)_{N_\tot}=0   $.
This result is in agreement with Ref.~\citenum{CohenMoriSYang08}.

\noindent \emph{2. Frontier eigenvalues and derivative discontinuity.}
We denote $M_B=(1-\alpha)S_0+\alpha S_1$, the highest value of $M_\tot$ (for a given $N_\tot)$, at the \emph{boundary} of Region~\eqref{eq:spin_condition}, and  consider $|M_\tot|<M_B$. 
In this case, 
$\lp( \de E_\ens / \de N_{\up} \rp)_{N_\dn} = \lp( \de E_\ens / \de N_{\dn} \rp)_{N_\up} = 
\lp( \de E_\ens / \de N_\tot \rp)_{M_\tot}$, because 
$\lp( \de E_\ens / \de M_\tot \rp)_{N_\tot} = 0$.  Hence, by Janak's theorem~\cite{Janak78}, the highest occupied (ho) KS energy eigenvalues are the same for both spin channels, constant strictly within each trapezoid/triangle as in Fig.~\ref{fig:Nup_Ndw}, and equal (the negative of) the corresponding ionization potential: $\eps_\ho^\up = \eps_\ho^\dn = E(N_0+1)-E(N_0)$.

However, at and beyond the boundary of Region~\eqref{eq:spin_condition}, the value of $\eps^\s_\ho$ may differ from $E(N_0+1)-E(N_0)$, as it is determined by the slope of the energy just outside and adjacently to the boundary. Whereas the full discussion of the energy profile outside Region~\eqref{eq:spin_condition} is beyond the
scope of this work, we note that just outside the boundary the energy is a plane determined by three pure-state energy values, one of which is outside Region~\eqref{eq:spin_condition}. Therefore, a change in the energy slope is expected as we cross the boundary $M_\tot = M_B$ of Region~\eqref{eq:spin_condition}.
Specifically, for \emph{spin migration} (variation of $M_\tot$ at constant $N_\tot$), we predict a \emph{derivative discontinuity} for the KS energy levels, which is manifested in a uniform jump in the KS potential $v_\KS^\s(\rr)$ at $M_\tot=M_B$ [a similar logic applies for the boundary $M_\tot = -M_B$].

We define the \emph{spin-migration derivative discontinuity} around $M_\tot = M_B$ as
\begin{align}\label{eq:def_Delta_sm}
&\nonumber \Delta_\textrm{sm}^\s =v_\KS^\s(\rr; M_B +\delta) - v_\KS^\s(\rr; M_B-\delta) 
\\& =\eps_i^\s(M_B + \delta)-\eps_i^\s(M_B-\delta)
\end{align}
where $\delta\to0^+$, $N_\tot$ is kept constant and $i$ refers to any of the KS energy levels.
In the following, quantities evaluated at $M_B+\delta$ are denoted with a prime (e.g.\ ${\eps'}_\ho^\s$) and those evaluated within Region~\eqref{eq:spin_condition} are without a prime.

As an illustration, choose 
$N_\tot\notin\mathbb{N}$ and consider a common example, where immediately outside Region~\eqref{eq:spin_condition}, for $M_\tot > M_B$, the ensemble ground state consists of $\ket{\Psi_{N_0,S_0}}$, $\ket{\Psi_{N_0,S_0+1}}$ and $\ket{\Psi_{N_0+1,S_1}}$.
Then, the ensemble energy there is $E'_\ens(N_\tot,M_\tot) = K N_\tot + J M_\tot + C$, where $J = E(N_0,S_0+1) - E(N_0,S_0)$ -- a spin-flip energy, and $K = (E(N_0+1) - E(N_0)) - (S_1-S_0) J$. Therefore, 
the slope changes around $M_B$ equal 
\begin{align}
\lp( \frac{\de E'_\ens}{\de N_\up} \rp)_{N_\dn} - \lp( \frac{\de E_\ens}{\de N_\up} \rp)_{N_\dn} = \lp(S_1 - S_0 - \half \rp) J \\
\lp( \frac{\de E'_\ens}{\de N_\dn} \rp)_{N_\up} - \lp( \frac{\de E_\ens}{\de N_\dn} \rp)_{N_\up} = \lp(S_1 - S_0 + \half \rp) J.
\end{align}
Notably, the $\up$-slope is continuous for $S_1 = S_0 + \thalf$ and the $\dn$-slope is continuous for $S_1 = S_0 - \thalf$. 

Specifically for $S_1 = S_0 + \thalf$, we obtain a discontinuity only in the $\dn$-channel:
\begin{equation} \label{eq:Delta_sm}
    \Delta_\textrm{sm}^\dn = J - \lp( {\eps'}_\ho ^\dn - {\eps'}_\lu ^\dn \rp).
\end{equation}
We note that as $M_\tot$ surpasses $M_B$, $N_\dn$ crosses an integer, while $N_\up$ does not; the $\dn$-ho orbital within Region~\eqref{eq:spin_condition} is the $\dn$-lu orbital immediately outside it.
The term in the parentheses in Eq.~\eqref{eq:Delta_sm} is (the negative of) the KS gap immediately outside Region~\eqref{eq:spin_condition}. It equals $\eps_{\ho-1}^\dn-\eps_{\ho}^\dn$ in terms of quantities inside Region~\eqref{eq:spin_condition}. 

In the uncommon case of $S_1 > S_0 + \thalf$, we expect discontinuities in both spin channels, while neither $N_\up$ nor $N_\dn$ cross an integer.

As to the value of the ho level at the boundary itself, it is determined by recalling that all the derivatives with respect to $N_\s$ are taken from below. Then, from geometric considerations in the $N_\up-N_\dn$ plane, it follows that $\eps_\ho^\up(M_B) = E(N_0+1) - E(N_0)$ if $S_1 \leqs S_0 + \thalf $ and  $\eps_\ho^\dn(M_B) = E(N_0+1) - E(N_0)$ if $S_1 \leqs S_0 - \thalf$. In other words, in the above cases the ho level at the boundary has the same value as strictly within~\eqref{eq:spin_condition}. Otherwise, at the boundary it has the value immediately outside~\eqref{eq:spin_condition}.

$\Delta_\textrm{sm}^\s$ is related to the xc correction to the spin stiffness defined in Ref.~\citenum{CapVigUll10}, for the case of integer $N_\tot$: if immediately outside Region~\eqref{eq:spin_condition} the ensemble ground state consists of $\ket{\Psi_{N_0,S_0}}$ and $\ket{\Psi_{N_0,S_0+1}}$, the two quantities coincide. However, in the general case they differ, as Ref.~\citenum{CapVigUll10} expressly makes use of excited-state spin ensembles, following Ref.~\citenum{Levy95}, and here we treat ground-state ensembles.

\noindent \emph{3. Non-uniqueness and degeneracy.}
Remarkably, the coefficients $\gamma_{N,M}$ of Eq.~\eqref{gamma} are confined by only \emph{three} constraints, Eq.~\eqref{eq:gamma_coeffs}, whereas the number of coefficients is $2(S_0+S_1+1)$. This means that the ground state $\hat{\Gamma}$ may not be uniquely defined. 
Strictly within Region~\eqref{eq:spin_condition} (i.e.\ $\abs{M_\tot}<M_B$) and $N_\tot\notin\mathbb{N}$, all ${2(S_0+S_1+1)}$ coefficients may be nonzero. Then, since there are 3 constraints (Eq.~\eqref{eq:gamma_coeffs}),  the degeneracy of the ground-state is ${(2S_0+2S_1-1)}$-fold. Similarly, if $N_\tot = N_0 \in\mathbb{N}$ and $\abs{M_\tot}<S_0$, then we are left with ${2S_0+1}$ coefficients and 2 constraints, hence the degeneracy is ${(2S_0-1)}$-fold.
We discuss this non-uniqueness here in detail, starting with the cases where $\hat{\Gamma}$ is defined uniquely.

\textit{Case (a)}. If $S_0=0$ and $S_1=\frac{1}{2}$ (e.g., adding an electron to H$^+$ towards H, while varying $M_\tot$), we are left with three uniquely determined $\gamma_{N,M}$'s: $\gamma_{N_0,0} = 1-\alpha$, $\gamma_{N_0+1,\frac{1}{2}} = \tfrac{1}{2}\alpha +  M_\tot$ and $\gamma_{N_0+1,-\frac{1}{2}} = \tfrac{1}{2}\alpha - M_\tot$. Similarly, if $S_0 = \tfrac{1}{2}$ and $S_1=0$ (e.g., adding an electron to H towards H$^-$, while varying $M_\tot$), we also have three uniquely defined coefficients: $\gamma_{N_0,\frac{1}{2}} = \frac{1}{2} (1-\alpha) + M_\tot$, $\gamma_{N_0,-\frac{1}{2}} = \frac{1}{2} (1-\alpha) - M_\tot$ and $\gamma_{N_0+1,0}=\alpha$. 

\textit{Case (b)}. At the boundary of Region~\eqref{eq:spin_condition}, where the spin is at its maximal value,  $M_\tot=M_B$, we find that the expression for $M_\tot$, $\sum_{M=-S_0}^{S_0} M \gamma_{N_0,M} + \sum_{M=-S_1}^{S_1} M \gamma_{N_0+1,M}$, reaches its maximum (under the aforementioned constraints on $\gamma_{N,M}$'s) only when  $\gamma_{N_0,S_0}=1-\alpha$,  $\gamma_{N_0+1, S_1}=\alpha$ and all the other $\gamma_{N,M}$'s vanish; the ground state is determined uniquely. If, for example, $S_1 = S_0 + \tfrac{1}{2}$, this corresponds to the common scenario of \emph{ionization} (via the up spin channel): fractionally varying $N_\up$ while $N_\dw$ is a constant integer. A unique solution is obtained also when $M_\tot$ reaches its minimal value, i.e., for $M_\tot=-M_B$.

\textit{Case (c).} For the scenario of \textit{spin migration}, when $N_\tot=N_0=\const.$ and $M_\tot$ varies between $-S_0$ and $S_0$, the ground state is \emph{not} defined uniquely, for $S_0 \geqs 1$. From Eq.~\eqref{eq:gamma_coeffs} we see that $\sum_{M=-S_1}^{S_1} \gamma_{N_0+1,M} = 0$ and therefore $\gamma_{N_0+1,M} = 0$, for all $M$. We are then left with $2S_0+1$ coefficients. 

In particular, for $S_0=1$ (e.g., spin migration for the C atom), the three coefficients $\gamma_{N_0,1}$, $\gamma_{N_0,0}$ and $\gamma_{N_0,-1}$ can be expressed as
$\gamma_{N_0,1} = \thalf x + \thalf M_\tot$, $\gamma_{N_0,0} = 1-x$ and $\gamma_{N_0,-1} = \thalf x - \thalf M_\tot$, where $x$ remains \emph{undetermined}. 
To understand the meaning of the above ambiguity in the ground state, focus on the case $M_\tot=0$. Then, $\hat{\Gamma} = \frac{x}{2} \ketbra{\Psi_{N_0,1}} + (1-x) \ketbra{\Psi_{N_0,0}} + \frac{x}{2} \ketbra{\Psi_{N_0,-1}}$,
meaning that to get the average $\statop{S_z} = 0$, the system must have an equal probability of $\tfrac{x}{2}$ to be in the $M=1$ and $M=-1$ states, and a complementary probability of $(1-x)$ to be in the state with $M=0$. Here we see a clear distinction of constraining the \emph{average} spin of the system being~0 versus requiring each and every replica in a macroscopic statistical ensemble to have a spin of~0. The latter is obtained by demanding that the ground state of the system is an eigenstate of $\hat{S}_z$, with eigenvalue 0 (which is equivalent here to setting $x=0$).

Surprisingly, despite the ambiguity in $\hat{\Gamma}$, in this Case the density, as well as the spin-densities, are determined unambiguously and do not depend on $x$. Since the pure-state densities $n_{N,M}(\rr)$ do not depend on $M$ (see the SI for details), $n_\ens(\rr)=n_{N_0}(\rr)$. 
 For the spin-densities,  $n^\s_{\ens}(\rr) = \half n_{N_0}(\rr) + \half \delta_\s Q_{\ens}(\rr)$, where $\delta_{\up} = 1$ and $\delta_{\dw} = -1$. The spin distribution, defined by
 $Q(\rr) \defeq n_\up (\rr) - n_\dw(\rr)$, is expressed for our ensemble as  
\begin{equation} \label{eq:Q_r_c}
 Q_\ens(\rr) =  M_\tot Q_{N_0,1}(\rr) = M_\tot (n^\up_{N_0,1}(\rr) - n^\dw_{N_0,1}(\rr)),
\end{equation}
being linear in $M_\tot$, and independent of $x$.
Here we used the fact that for any pure state $\ket{\Psi_{N,M}}$, $n^\up_{N,-M}(\rr) = n^\dw_{N,M}(\rr)$. 

\textit{Case (d).} For spin migration with $S_0 = \tfrac{3}{2}$ (e.g., for the N atom), even the spin-densities are \emph{not} determined uniquely anymore. Here we have four coefficients (dropping the index $N_0$ for brevity): $\gamma_{3/2} = \thalf x + \tfrac{1}{3}(M_\tot - y)$, $\gamma_{1/2} = \thalf (1-x) + y$, $\gamma_{-1/2} = \thalf (1-x) - y$ and $\gamma_{-3/2} = \thalf x - \tfrac{1}{3}(M_\tot - y)$, with two undetermined parameters, $x$ and $y$. Whereas the total energy and the total density are determined unambiguously, the spin-densities linearly depend on $y$ and are independent of~$x$. The ensemble spin distribution
\begin{align} \label{eq:Q_r_d}
Q_\ens(\rr) = (M_\tot-y)\frac{Q_{3/2}(\rr)}{3/2}  + y \frac{Q_{1/2}(\rr)}{1/2} 
\end{align}
is linear in $y$.

\textit{Case (e).} For a fractional $N_\tot=N_0+\alpha$, $S_0=\frac{1}{2}$ and $S_1=1$ (e.g., adding an electron to C$^+$ towards C, while varying $M_\tot$), we end up with five coefficients, which depend on two parameters, $x$ and $y$: $\gamma_{N_0,1/2} = \thalf (1-\alpha) + y$, $\gamma_{N_0,-1/2} = \thalf (1-\alpha) - y$, $\gamma_{N_0+1,1} = \thalf (\alpha-x) + \thalf (M_\tot - y)$, $\gamma_{N_0+1,0} = x$ and $\gamma_{N_0+1,-1} = \thalf (\alpha-x) - \thalf (M_\tot - y)$. The total ensemble density $n_\ens(\rr) = (1-\alpha) n_{N_0}(\rr) + \alpha n_{N_0+1}(\rr)$ is piecewise-linear in $\alpha$ and remains fully determined, whereas the spin-densities are not. The ensemble spin distribution is
\begin{equation} \label{eq:Q_r_e}
        Q_\ens(\rr) = (M_\tot-y) Q_{N_0+1,1}(\rr) + y \frac{Q_{N_0,1/2}(\rr)}{1/2},
\end{equation}
being linear in $y$, but independent of $\alpha$ (in full contrast to $n_\ens(\rr)$). 

Cases (d) and (e) clearly show that, for a given electron number and spin, we have a set of degenerate ground states $\hat{\Gamma}(x,y)$, with the \emph{same} total density, but with different spin-densities, $n^\up_\ens(\rr)$ and $n^\dn_\ens(\rr)$. Notably, even in the spin-unpolarized case $M_\tot=0$, the spin-densities are not determined, and are not necessarily equal to each other.

\noindent \emph{4. Removing non-uniqueness}. To further explore and interpret the surprising result of non-uniqueness at the ground state, we offer two ways for its full or partial removal. 

First, to determine the ground state unambiguously, one can introduce additional constraints, to complement~\eqref{constraint_N}--\eqref{constraint_prob}. For example, we can set $\stat{\hat{S}_z^2}$, $\stat{\hat{S}_z^3}$ and/or higher moments of $\hat{S}_z$.  In the SI we analyze how many such moments are sufficient to determine $\hat{\Gamma}$ and/or $Q_\ens(\rr)$, in the general case.
Specifically, for Case~(c) above, $\stat{\hat{S}_z^2} = x$, i.e., setting $\stat{\hat{S}_z^2}$ fully determines $\hat{\Gamma}$. 

Alternatively, one can require the standard deviation $\Delta S_z \defeq \sqrt{\stat{\hat{S}_z^2} - \stat{\hat{S}_z}^2}$ to be minimal. 
This requirement sets the ground state \emph{uniquely}; a mathematical proof of this statement is provided in the SI.
In particular, in Case (c) the deviation is $\Delta S_z = \sqrt{x-M_\tot^2}$. For all coefficients $\gamma_{N,M}$ to remain within $[0,1]$, $x$ is confined to $[|M_\tot|,1]$, and therefore $(\Delta S_z)_\m = \sqrt{|M_\tot| - M_\tot^2}$.  Notably, for minimal $\Delta S_z$ we are left with only two non-zero coefficients: For $M_\tot \in [0,1]$, $\gamma_{N_0,1}=M_\tot$ and $\gamma_{N_0,0} = 1 - M_\tot$; for $M_\tot \in [-1,0]$, $\gamma_{N_0,-1}=-M_\tot$ and $\gamma_{N_0,0} = 1 + M_\tot$.  Thus, as $M_\tot$ goes from 1 to -1, we observe a gradual transition between $\ketbra{\Psi_{N_0,1}}$ to $\ketbra{\Psi_{N_0,0}}$ and then to $\ketbra{\Psi_{N_0,-1}}$, with two adjacent pure states involved at each point. 

In Case (d), $\stat{\hat{S}_z^2}$, as well as all even moments of $\hat{S}_z$ equal $\stat{\hat{S}_z^{2k}} = (S_0-1)^{2k} (1-x) + S_0^{2k} x$, being linear in $x$ and independent of $y$ ($k \in \mathbb{N}$). Conversely, all the odd moments $\stat{\hat{S}_z^{2k+1}} = (S_0-1)^{2k} y + S_0^{2k} (M_\tot - y)$ are linear in $y$ and independent of $x$.  Therefore, it is sufficient to set $\stat{\hat{S}_z^2}$ and $\stat{\hat{S}_z^3}$,  or to minimize $\Delta S_z$ (see the SI for the technical details). 
In the latter case, we obtain a gradual transition from  $\ketbra{\Psi_{N_0,3/2}}$ to $\ketbra{\Psi_{N_0,1/2}}$ to $\ketbra{\Psi_{N_0,-1/2}}$ to $\ketbra{\Psi_{N_0,-3/2}}$, with two adjacent pure states involved at each point. The spin distribution is unambiguously determined and piecewise-linear in $M_\tot$:
\begin{align}
\label{eq:zeta_min_DSz}
& Q_\ens(\rr)= \nonumber \\
 &\begin{cases}
			-Q_{1/2}(\rr) + (M_\tot \!+\! \thalf) (Q_{3/2}(\rr) \!-\! Q_{1/2}(\rr))
            & \!\!\!\! : M_\tot \in [-\tfrac{3}{2}, -\thalf]\\
           2 M_\tot Q_{1/2}(\rr) &
            \!\!\!\! : M_\tot \in [-\thalf, \thalf]\\ 
            Q_{1/2}(\rr) + (M_\tot - \thalf) (Q_{3/2}(\rr) - Q_{1/2}(\rr)) &\!\!\!\! : M_\tot \in [\thalf, \tfrac{3}{2}]
		 \end{cases}
\end{align}

Second, to remove non-uniqueness of the ground state it is natural to introduce a weak magnetic field~\footnote{The magnetic field being weak means that $\muB B_0$ is smaller than any energy difference in the problem; $f(\rr)$ is bounded. Therefore, we refer only to terms that are first order in $\mathbf{B}$.}.
Surprisingly, while lifting the degeneracy between pure states $\ket{\Psi_{N,M}}$ with different $M$, a magnetic field does not fully remove the non-uniqueness.

Consider an inhomogeneous magnetic field $\mathbf{B}(\rr) = B_0 f(\rr) \hat{z}$. 
Then, the pure-state energies become $\tilde{E}(N,M) = E(N) + \muB B_0 \int f(\rr) (n^\up_{N,M}(\rr)-n^\dw_{N,M}(\rr)) \, d^3r = E(N) + \muB B_0 M F_{N,M}$, where $F_{N,M}= \tfrac{1}{M} \int f(\rr)  (n^\up_{N,M}(\rr)-n^\dw_{N,M}(\rr)) \, d^3r$ for $M \neq 0$ and 0 otherwise%
~\footnote{In our analysis we disregard, for simplicity, the magnetic term $\mathbf{A} \cdot \mathbf{\hat{p}}$, where $\mathbf{A}$ is the electromagnetic vector-potential and $\mathbf{\hat{p}}$ is the momentum operator. [For homogeneous magnetic fields, this term boils down to $\mathbf{B} \cdot \mathbf{\hat{L}}$]. Treatment of this term changes the value of the energy $E(N)$, but it is independent of $M$ -- which is our main focus.}.

Applying this magnetic field in Case (d), the ensemble energy becomes $\tilde{E}_\ens(N_0,M_\tot) = E(N_0) +  \muB B_0 M_\tot F_{N_0,3/2} +  \muB B_0 (F_{N_0,1/2}-F_{N_0,3/2}) y$, being linear in $y$. One can then minimize $\tilde{E}_\ens(N_0,M_\tot)$ with respect to $y$: 
If, say, for a given system and field profile  $F_{N_0,1/2} 
> F_{N_0,3/2}$, then for $B_0 > 0$, we seek for the lowest possible $y$ and for $B_0 < 0$ -- for the highest. The range of $y$ values is confined by the requirement $\gamma_{N,M} \in [0, 1]$. In this Case, it results in $x \in [0,1]$, $|y| \leqs \thalf (1-x)$ and $|y-M_\tot| \leqs \tfrac{3}{2} x$. Consequently, the lowest possible $y$ is $-\thalf M_\tot - \tfrac{3}{4}$, for $M_\tot \in [-\tfrac{3}{2}, -\thalf]$ and $\tfrac{1}{4} M_\tot - \tfrac{3}{8}$, for $M_\tot \in [-\thalf,\tfrac{3}{2}]$. Fortunately, minimizing $y$ in this Case also uniquely sets $x$, and therefore the ground state is determined uniquely.  As in the case of minimal $\Delta S_z$, also here we are left with only two pure states involved in the ensemble at each point. Now, however, we perform a gradual transition from  $\ketbra{\Psi_{N_0,3/2}}$ to $\ketbra{\Psi_{N_0,-1/2}}$ for $M_\tot : \tfrac{3}{2} \rarr -\thalf$, and then from $\ketbra{\Psi_{N_0,-1/2}}$ to $\ketbra{\Psi_{N_0,-3/2}}$, for $M_\tot : -\thalf \rarr -\tfrac{3}{2}$.

In contrast, in Case (c), the ensemble energy in presence of the magnetic field becomes $\tilde{E}_\ens(N_0,M_\tot) = E(N_0) + \muB B_0 M_\tot F_{N_0,1}$, being independent of $x$.  Therefore, the non-uniqueness of the ground state cannot be removed.

In Case (e), the ensemble energy becomes $\tilde{E}_\ens(N_\tot,M_\tot) = (1-\alpha) E(N_0) + \alpha E(N_0+1) +  \muB B_0 M_\tot F_{N_0,1} + \muB B_0 (F_{N_0,1/2}-F_{N_0+1,1}) y$, and can be minimized with respect to $y$. However, as opposed to Case (d), here a minimal/maximal value of $y$ can correspond to a range of possible $x$ values; the ground state is not fully determined. This is not surprising, as Case (e) has Case (c) as a particular limit, for $\alpha=1$ (see the SI for details).

Finally, we note that applying a strong magnetic field could further remove the ambiguity in the ground state. However, this scenario may change which pure states contribute to the ensemble ground state, and therefore it is outside the scope of this Letter.

To conclude, in this Letter we rigorously described the exact ground state of a finite many-electron system, with $N_\tot$ electrons and with spin $M_\tot$, fully employing the ensemble approach. Our results hold for any value of $N_\tot$, any values of the equilibrium spin, $S_{\m}(N)$, and for $M_\tot$ confined by Eq.~\eqref{eq:spin_condition}. Description outside of Region \eqref{eq:spin_condition} is a subject for future work. 

First, we found that the ground state is an ensemble, which is comprised of $2(S_0+S_1+1)$ pure states, as indicated in Eq.~\eqref{gamma}. These states are not necessarily the four integer neighboring points to $(N_\tot,M_\tot)$ on the $N_\up-N_\dn$ grid. The graph of the total energy, $E_\ens(N_\tot,M_\tot)$ consists therefore not only of triangles, but also of trapezoids, when $S_{\m} \geqs 1$. The total energy, the total density, and any quantity $A$, whose pure-state expectation values $\expval{\hat{A}}{\Psi_{N,M}}$ do not depend on $M$, are piecewise-linear in $N_\tot$ (i.e., in $\alpha$) and independent of $M_\tot$. 

Second, the highest-occupied KS $\up$- and $\dn$-orbital energies are found to be equal strictly within Region~\eqref{eq:spin_condition}, and the KS potentials experience a spatially uniform ``jump'' at the boundary of this region (Eq.~\eqref{eq:def_Delta_sm}), directly related to the spin-migration derivative discontinuity, $\Delta_\textrm{sm}^\s$ (Eq.~\eqref{eq:def_Delta_sm}), introduced here for the first time.

Third, we discovered that, generally speaking, the ground state $\hat{\Gamma}$ is not uniquely defined. Unlike the total energy and the total density, which are determined uniquely, the spin-densities are not (Eqs.~\eqref{eq:Q_r_d} and~\eqref{eq:Q_r_e}). Two ways of removing this non-uniqueness are suggested: adding constraints, e.g., on higher moments of $\hat{S}_z$, and minimizing $\Delta S_z$. Applying a weak magnetic field does \emph{not} always fully remove the non-uniqueness. 

Our findings generalize the piecewise-linearity and the flat-plane conditions and provide new exact properties for many-electron systems. These are expected to be useful in the development of advanced approximations in density functional theory and in other many-electron methods.
  
\section*{Acknowledgement} 
The authors thank Prof.~Roi Baer for illuminating comments and fruitful discussions.

\section*{Supporting Information Available}

In the SI we provide technical details regarding four subjects.
(i) We show that certain pure-state expectation values, including the density $\expval{\hat{n}(\rr)}{\Psi_{N,S,M}}$ do not depend on $M$. (ii) We discuss the uniqueness of $\hat{\Gamma}$ and $Q_\ens(\rr)$, given the values of certain moments of $\hat{S}_z$. (iii) We prove that $\hat{\Gamma}$ is uniquely determined by the added constraint of minimizing $\Delta S_z$. (iv) We elaborate on Cases (c), (d) and (e), regarding the spin-densities, magnetic fields, and the moments of $\hat{S}_z$.

\bibliography{bib_EK_2022}

\begin{thebibliography}{36}%
\makeatletter
\providecommand \@ifxundefined [1]{%
 \@ifx{#1\undefined}
}%
\providecommand \@ifnum [1]{%
 \ifnum #1\expandafter \@firstoftwo
 \else \expandafter \@secondoftwo
 \fi
}%
\providecommand \@ifx [1]{%
 \ifx #1\expandafter \@firstoftwo
 \else \expandafter \@secondoftwo
 \fi
}%
\providecommand \natexlab [1]{#1}%
\providecommand \enquote  [1]{``#1''}%
\providecommand \bibnamefont  [1]{#1}%
\providecommand \bibfnamefont [1]{#1}%
\providecommand \citenamefont [1]{#1}%
\providecommand \href@noop [0]{\@secondoftwo}%
\providecommand \href [0]{\begingroup \@sanitize@url \@href}%
\providecommand \@href[1]{\@@startlink{#1}\@@href}%
\providecommand \@@href[1]{\endgroup#1\@@endlink}%
\providecommand \@sanitize@url [0]{\catcode `\\12\catcode `\$12\catcode `\&12\catcode `\#12\catcode `\^12\catcode `\_12\catcode `\%12\relax}%
\providecommand \@@startlink[1]{}%
\providecommand \@@endlink[0]{}%
\providecommand \url  [0]{\begingroup\@sanitize@url \@url }%
\providecommand \@url [1]{\endgroup\@href {#1}{\urlprefix }}%
\providecommand \urlprefix  [0]{URL }%
\providecommand \Eprint [0]{\href }%
\providecommand \doibase [0]{https://doi.org/}%
\providecommand \selectlanguage [0]{\@gobble}%
\providecommand \bibinfo  [0]{\@secondoftwo}%
\providecommand \bibfield  [0]{\@secondoftwo}%
\providecommand \translation [1]{[#1]}%
\providecommand \BibitemOpen [0]{}%
\providecommand \bibitemStop [0]{}%
\providecommand \bibitemNoStop [0]{.\EOS\space}%
\providecommand \EOS [0]{\spacefactor3000\relax}%
\providecommand \BibitemShut  [1]{\csname bibitem#1\endcsname}%
\let\auto@bib@innerbib\@empty
\bibitem [{\citenamefont {Szabo}\ and\ \citenamefont {Ostlund}(1996)}]{Szabo}%
  \BibitemOpen
  \bibfield  {author} {\bibinfo {author} {\bibfnamefont {A.}~\bibnamefont {Szabo}}\ and\ \bibinfo {author} {\bibfnamefont {N.}~\bibnamefont {Ostlund}},\ }\href@noop {} {\emph {\bibinfo {title} {Modern Quantum Chemistry: Introduction to Advanced Electronic Structure Theory}}}\ (\bibinfo  {publisher} {Dover, Mineola},\ \bibinfo {year} {1996})\BibitemShut {NoStop}%
\bibitem [{\citenamefont {Parr}\ and\ \citenamefont {Yang}(1989)}]{PY}%
  \BibitemOpen
  \bibfield  {author} {\bibinfo {author} {\bibfnamefont {R.~G.}\ \bibnamefont {Parr}}\ and\ \bibinfo {author} {\bibfnamefont {W.}~\bibnamefont {Yang}},\ }\href@noop {} {\emph {\bibinfo {title} {{Density-Functional Theory of Atoms and Molecules}}}}\ (\bibinfo  {publisher} {Oxford University Press},\ \bibinfo {year} {1989})\BibitemShut {NoStop}%
\bibitem [{\citenamefont {{R.~M. Dreizler}}\ and\ \citenamefont {{E.~K.~U. Gross}}(1990)}]{DG}%
  \BibitemOpen
  \bibfield  {author} {\bibinfo {author} {\bibnamefont {{R.~M. Dreizler}}}\ and\ \bibinfo {author} {\bibnamefont {{E.~K.~U. Gross}}},\ }\href@noop {} {\emph {\bibinfo {title} {{D}ensity {F}unctional {T}heory}}}\ (\bibinfo  {publisher} {Springer Verlag},\ \bibinfo {year} {1990})\BibitemShut {NoStop}%
\bibitem [{\citenamefont {Fiolhais}\ \emph {et~al.}(2003)\citenamefont {Fiolhais}, \citenamefont {Nogueira},\ and\ \citenamefont {Marques}}]{Primer}%
  \BibitemOpen
  \bibinfo {editor} {\bibfnamefont {C.}~\bibnamefont {Fiolhais}}, \bibinfo {editor} {\bibfnamefont {F.}~\bibnamefont {Nogueira}},\ and\ \bibinfo {editor} {\bibfnamefont {M.~A.~L.}\ \bibnamefont {Marques}},\ eds.,\ \href@noop {} {\emph {\bibinfo {title} {A {P}rimer in {D}ensity {F}unctional {T}heory}}}\ (\bibinfo  {publisher} {Springer},\ \bibinfo {year} {2003})\BibitemShut {NoStop}%
\bibitem [{\citenamefont {Fetter}\ and\ \citenamefont {Walecka}(1971)}]{FetterWalecka}%
  \BibitemOpen
  \bibfield  {author} {\bibinfo {author} {\bibfnamefont {A.~L.}\ \bibnamefont {Fetter}}\ and\ \bibinfo {author} {\bibfnamefont {J.~D.}\ \bibnamefont {Walecka}},\ }\href@noop {} {\emph {\bibinfo {title} {{Quantum Theory of Many-Particle Systems}}}}\ (\bibinfo  {publisher} {McGraw-Hill},\ \bibinfo {address} {New York},\ \bibinfo {year} {1971})\BibitemShut {NoStop}%
\bibitem [{\citenamefont {Martin}\ \emph {et~al.}(2016)\citenamefont {Martin}, \citenamefont {Reining},\ and\ \citenamefont {Ceperley}}]{Reining_MB}%
  \BibitemOpen
  \bibfield  {author} {\bibinfo {author} {\bibfnamefont {R.~M.}\ \bibnamefont {Martin}}, \bibinfo {author} {\bibfnamefont {L.}~\bibnamefont {Reining}},\ and\ \bibinfo {author} {\bibfnamefont {D.~M.}\ \bibnamefont {Ceperley}},\ }\href@noop {} {\emph {\bibinfo {title} {{I}nteracting {E}lectrons: {T}heory and {C}omputational {A}pproaches}}}\ (\bibinfo  {publisher} {Cambridge University Press},\ \bibinfo {year} {2016})\BibitemShut {NoStop}%
\bibitem [{\citenamefont {Perdew}\ \emph {et~al.}(2005)\citenamefont {Perdew}, \citenamefont {Ruzsinszky}, \citenamefont {Tao}, \citenamefont {Staroverov}, \citenamefont {Scuseria},\ and\ \citenamefont {Csonka}}]{Perdew05}%
  \BibitemOpen
  \bibfield  {author} {\bibinfo {author} {\bibfnamefont {J.~P.}\ \bibnamefont {Perdew}}, \bibinfo {author} {\bibfnamefont {A.}~\bibnamefont {Ruzsinszky}}, \bibinfo {author} {\bibfnamefont {J.}~\bibnamefont {Tao}}, \bibinfo {author} {\bibfnamefont {V.~N.}\ \bibnamefont {Staroverov}}, \bibinfo {author} {\bibfnamefont {G.~E.}\ \bibnamefont {Scuseria}},\ and\ \bibinfo {author} {\bibfnamefont {G.~I.}\ \bibnamefont {Csonka}},\ }\href {https://doi.org/10.1063/1.1904565} {\bibfield  {journal} {\bibinfo  {journal} {The Journal of Chemical Physics}\ }\textbf {\bibinfo {volume} {123}},\ \bibinfo {pages} {062201} (\bibinfo {year} {2005})},\ \Eprint {https://arxiv.org/abs/https://pubs.aip.org/aip/jcp/article-pdf/doi/10.1063/1.1904565/15370901/062201\_1\_online.pdf} {https://pubs.aip.org/aip/jcp/article-pdf/doi/10.1063/1.1904565/15370901/062201\_1\_online.pdf} \BibitemShut {NoStop}%
\bibitem [{\citenamefont {Perdew}\ \emph {et~al.}(2009)\citenamefont {Perdew}, \citenamefont {Ruzsinszky}, \citenamefont {Constantin}, \citenamefont {Sun},\ and\ \citenamefont {Csonka}}]{Perdew09_perplexed}%
  \BibitemOpen
  \bibfield  {author} {\bibinfo {author} {\bibfnamefont {J.~P.}\ \bibnamefont {Perdew}}, \bibinfo {author} {\bibfnamefont {A.}~\bibnamefont {Ruzsinszky}}, \bibinfo {author} {\bibfnamefont {L.~A.}\ \bibnamefont {Constantin}}, \bibinfo {author} {\bibfnamefont {J.}~\bibnamefont {Sun}},\ and\ \bibinfo {author} {\bibfnamefont {G.~I.}\ \bibnamefont {Csonka}},\ }\href@noop {} {\bibfield  {journal} {\bibinfo  {journal} {J. Chem. Theory Comp.}\ }\textbf {\bibinfo {volume} {5}},\ \bibinfo {pages} {902} (\bibinfo {year} {2009})}\BibitemShut {NoStop}%
\bibitem [{\citenamefont {Kaplan}\ \emph {et~al.}(2023)\citenamefont {Kaplan}, \citenamefont {Levy},\ and\ \citenamefont {Perdew}}]{KaplanLevyPerdew23}%
  \BibitemOpen
  \bibfield  {author} {\bibinfo {author} {\bibfnamefont {A.~D.}\ \bibnamefont {Kaplan}}, \bibinfo {author} {\bibfnamefont {M.}~\bibnamefont {Levy}},\ and\ \bibinfo {author} {\bibfnamefont {J.~P.}\ \bibnamefont {Perdew}},\ }\href@noop {} {\bibfield  {journal} {\bibinfo  {journal} {Annu. Rev. Phys. Chem.}\ }\textbf {\bibinfo {volume} {74}},\ \bibinfo {pages} {193} (\bibinfo {year} {2023})}\BibitemShut {NoStop}%
\bibitem [{\citenamefont {Perdew}\ \emph {et~al.}(1982)\citenamefont {Perdew}, \citenamefont {Parr}, \citenamefont {Levy},\ and\ \citenamefont {Balduz}}]{PPLB82}%
  \BibitemOpen
  \bibfield  {author} {\bibinfo {author} {\bibfnamefont {J.~P.}\ \bibnamefont {Perdew}}, \bibinfo {author} {\bibfnamefont {R.~G.}\ \bibnamefont {Parr}}, \bibinfo {author} {\bibfnamefont {M.}~\bibnamefont {Levy}},\ and\ \bibinfo {author} {\bibfnamefont {J.~L.}\ \bibnamefont {Balduz}},\ }\href@noop {} {\bibfield  {journal} {\bibinfo  {journal} {Phys. Rev. Lett.}\ }\textbf {\bibinfo {volume} {49}},\ \bibinfo {pages} {1691} (\bibinfo {year} {1982})}\BibitemShut {NoStop}%
\bibitem [{\citenamefont {Heinrich}\ \emph {et~al.}(2004)\citenamefont {Heinrich}, \citenamefont {Gupta}, \citenamefont {Lutz},\ and\ \citenamefont {Eigler}}]{Heinrich04}%
  \BibitemOpen
  \bibfield  {author} {\bibinfo {author} {\bibfnamefont {A.~J.}\ \bibnamefont {Heinrich}}, \bibinfo {author} {\bibfnamefont {J.~A.}\ \bibnamefont {Gupta}}, \bibinfo {author} {\bibfnamefont {C.~P.}\ \bibnamefont {Lutz}},\ and\ \bibinfo {author} {\bibfnamefont {D.~M.}\ \bibnamefont {Eigler}},\ }\href {https://doi.org/10.1126/science.1101077} {\bibfield  {journal} {\bibinfo  {journal} {Science}\ }\textbf {\bibinfo {volume} {306}},\ \bibinfo {pages} {466} (\bibinfo {year} {2004})},\ \Eprint {https://arxiv.org/abs/https://www.science.org/doi/pdf/10.1126/science.1101077} {https://www.science.org/doi/pdf/10.1126/science.1101077} \BibitemShut {NoStop}%
\bibitem [{\citenamefont {Capelle}\ \emph {et~al.}(2010)\citenamefont {Capelle}, \citenamefont {Vignale},\ and\ \citenamefont {Ullrich}}]{CapVigUll10}%
  \BibitemOpen
  \bibfield  {author} {\bibinfo {author} {\bibfnamefont {K.}~\bibnamefont {Capelle}}, \bibinfo {author} {\bibfnamefont {G.}~\bibnamefont {Vignale}},\ and\ \bibinfo {author} {\bibfnamefont {C.~A.}\ \bibnamefont {Ullrich}},\ }\href@noop {} {\bibfield  {journal} {\bibinfo  {journal} {Phys. Rev. B}\ }\textbf {\bibinfo {volume} {81}},\ \bibinfo {pages} {125114} (\bibinfo {year} {2010})}\BibitemShut {NoStop}%
\bibitem [{\citenamefont {Casanova}\ and\ \citenamefont {Krylov}(2020)}]{CasanovaKrylov20}%
  \BibitemOpen
  \bibfield  {author} {\bibinfo {author} {\bibfnamefont {D.}~\bibnamefont {Casanova}}\ and\ \bibinfo {author} {\bibfnamefont {A.~I.}\ \bibnamefont {Krylov}},\ }\href {https://doi.org/10.1039/C9CP06507E} {\bibfield  {journal} {\bibinfo  {journal} {Phys. Chem. Chem. Phys.}\ }\textbf {\bibinfo {volume} {22}},\ \bibinfo {pages} {4326} (\bibinfo {year} {2020})}\BibitemShut {NoStop}%
\bibitem [{\citenamefont {Chan}(1999)}]{Chan99}%
  \BibitemOpen
  \bibfield  {author} {\bibinfo {author} {\bibfnamefont {G.~K.-L.}\ \bibnamefont {Chan}},\ }\href@noop {} {\bibfield  {journal} {\bibinfo  {journal} {J. Chem. Phys.}\ }\textbf {\bibinfo {volume} {110}},\ \bibinfo {pages} {4710} (\bibinfo {year} {1999})}\BibitemShut {NoStop}%
\bibitem [{\citenamefont {Cohen}\ \emph {et~al.}(2008{\natexlab{a}})\citenamefont {Cohen}, \citenamefont {Mori-Sánchez},\ and\ \citenamefont {Yang}}]{CohenMoriSYang08}%
  \BibitemOpen
  \bibfield  {author} {\bibinfo {author} {\bibfnamefont {A.~J.}\ \bibnamefont {Cohen}}, \bibinfo {author} {\bibfnamefont {P.}~\bibnamefont {Mori-Sánchez}},\ and\ \bibinfo {author} {\bibfnamefont {W.}~\bibnamefont {Yang}},\ }\href {https://doi.org/10.1063/1.2987202} {\bibfield  {journal} {\bibinfo  {journal} {The Journal of Chemical Physics}\ }\textbf {\bibinfo {volume} {129}},\ \bibinfo {pages} {121104} (\bibinfo {year} {2008}{\natexlab{a}})},\ \Eprint {https://arxiv.org/abs/https://pubs.aip.org/aip/jcp/article-pdf/doi/10.1063/1.2987202/15419668/121104\_1\_online.pdf} {https://pubs.aip.org/aip/jcp/article-pdf/doi/10.1063/1.2987202/15419668/121104\_1\_online.pdf} \BibitemShut {NoStop}%
\bibitem [{\citenamefont {Mori-S\'anchez}\ \emph {et~al.}(2009)\citenamefont {Mori-S\'anchez}, \citenamefont {Cohen},\ and\ \citenamefont {Yang}}]{MoriS09}%
  \BibitemOpen
  \bibfield  {author} {\bibinfo {author} {\bibfnamefont {P.}~\bibnamefont {Mori-S\'anchez}}, \bibinfo {author} {\bibfnamefont {A.~J.}\ \bibnamefont {Cohen}},\ and\ \bibinfo {author} {\bibfnamefont {W.}~\bibnamefont {Yang}},\ }\href {https://doi.org/10.1103/PhysRevLett.102.066403} {\bibfield  {journal} {\bibinfo  {journal} {Phys. Rev. Lett.}\ }\textbf {\bibinfo {volume} {102}},\ \bibinfo {pages} {066403} (\bibinfo {year} {2009})}\BibitemShut {NoStop}%
\bibitem [{\citenamefont {Gál}\ and\ \citenamefont {Geerlings}(2010)}]{GalGeerlings10JCP}%
  \BibitemOpen
  \bibfield  {author} {\bibinfo {author} {\bibfnamefont {T.}~\bibnamefont {Gál}}\ and\ \bibinfo {author} {\bibfnamefont {P.}~\bibnamefont {Geerlings}},\ }\href {https://doi.org/10.1063/1.3467898} {\bibfield  {journal} {\bibinfo  {journal} {The Journal of Chemical Physics}\ }\textbf {\bibinfo {volume} {133}},\ \bibinfo {pages} {144105} (\bibinfo {year} {2010})},\ \Eprint {https://arxiv.org/abs/https://pubs.aip.org/aip/jcp/article-pdf/doi/10.1063/1.3467898/15432041/144105\_1\_online.pdf} {https://pubs.aip.org/aip/jcp/article-pdf/doi/10.1063/1.3467898/15432041/144105\_1\_online.pdf} \BibitemShut {NoStop}%
\bibitem [{\citenamefont {G\'al}\ and\ \citenamefont {Geerlings}(2010)}]{GalGeerlings10PRA}%
  \BibitemOpen
  \bibfield  {author} {\bibinfo {author} {\bibfnamefont {T.}~\bibnamefont {G\'al}}\ and\ \bibinfo {author} {\bibfnamefont {P.}~\bibnamefont {Geerlings}},\ }\href {https://doi.org/10.1103/PhysRevA.81.032512} {\bibfield  {journal} {\bibinfo  {journal} {Phys. Rev. A}\ }\textbf {\bibinfo {volume} {81}},\ \bibinfo {pages} {032512} (\bibinfo {year} {2010})}\BibitemShut {NoStop}%
\bibitem [{\citenamefont {Cohen}\ \emph {et~al.}(2012)\citenamefont {Cohen}, \citenamefont {Mori-S\'{a}nchez},\ and\ \citenamefont {Yang}}]{cohen12}%
  \BibitemOpen
  \bibfield  {author} {\bibinfo {author} {\bibfnamefont {A.~J.}\ \bibnamefont {Cohen}}, \bibinfo {author} {\bibfnamefont {P.}~\bibnamefont {Mori-S\'{a}nchez}},\ and\ \bibinfo {author} {\bibfnamefont {W.}~\bibnamefont {Yang}},\ }\href@noop {} {\bibfield  {journal} {\bibinfo  {journal} {Chem. Rev.}\ }\textbf {\bibinfo {volume} {112}},\ \bibinfo {pages} {289} (\bibinfo {year} {2012})}\BibitemShut {NoStop}%
\bibitem [{\citenamefont {Cuevas-Saavedra}\ \emph {et~al.}(2012)\citenamefont {Cuevas-Saavedra}, \citenamefont {Chakraborty}, \citenamefont {Rabi}, \citenamefont {C{\'a}rdenas},\ and\ \citenamefont {Ayers}}]{Saavedra12}%
  \BibitemOpen
  \bibfield  {author} {\bibinfo {author} {\bibfnamefont {R.}~\bibnamefont {Cuevas-Saavedra}}, \bibinfo {author} {\bibfnamefont {D.}~\bibnamefont {Chakraborty}}, \bibinfo {author} {\bibfnamefont {S.}~\bibnamefont {Rabi}}, \bibinfo {author} {\bibfnamefont {C.}~\bibnamefont {C{\'a}rdenas}},\ and\ \bibinfo {author} {\bibfnamefont {P.~W.}\ \bibnamefont {Ayers}},\ }\href@noop {} {\bibfield  {journal} {\bibinfo  {journal} {J. Chem. Theory Comput.}\ }\textbf {\bibinfo {volume} {8}},\ \bibinfo {pages} {4081} (\bibinfo {year} {2012})}\BibitemShut {NoStop}%
\bibitem [{\citenamefont {Gould}\ and\ \citenamefont {Dobson}(2013)}]{GouldDobson13}%
  \BibitemOpen
  \bibfield  {author} {\bibinfo {author} {\bibfnamefont {T.}~\bibnamefont {Gould}}\ and\ \bibinfo {author} {\bibfnamefont {J.~F.}\ \bibnamefont {Dobson}},\ }\href@noop {} {\bibfield  {journal} {\bibinfo  {journal} {J. Chem. Phys.}\ }\textbf {\bibinfo {volume} {138}},\ \bibinfo {pages} {014103} (\bibinfo {year} {2013})}\BibitemShut {NoStop}%
\bibitem [{\citenamefont {Yang}\ \emph {et~al.}(2016)\citenamefont {Yang}, \citenamefont {Patel}, \citenamefont {Miranda-Quintana}, \citenamefont {Heidar-Zadeh}, \citenamefont {lez Espinoza},\ and\ \citenamefont {Ayers}}]{XDYang16}%
  \BibitemOpen
  \bibfield  {author} {\bibinfo {author} {\bibfnamefont {X.~D.}\ \bibnamefont {Yang}}, \bibinfo {author} {\bibfnamefont {A.~H.}\ \bibnamefont {Patel}}, \bibinfo {author} {\bibfnamefont {R.~A.}\ \bibnamefont {Miranda-Quintana}}, \bibinfo {author} {\bibfnamefont {F.}~\bibnamefont {Heidar-Zadeh}}, \bibinfo {author} {\bibfnamefont {C.~E.}\ \bibnamefont {lez Espinoza}},\ and\ \bibinfo {author} {\bibfnamefont {P.~W.}\ \bibnamefont {Ayers}},\ }\href@noop {} {\bibfield  {journal} {\bibinfo  {journal} {J Chem Phys}\ }\textbf {\bibinfo {volume} {145}},\ \bibinfo {pages} {031102} (\bibinfo {year} {2016})}\BibitemShut {NoStop}%
\bibitem [{Note1()}]{Note1}%
  \BibitemOpen
  \bibinfo {note} {Hartree atomic units are used throughout}\BibitemShut {NoStop}%
\bibitem [{\citenamefont {Cohen}\ \emph {et~al.}(2008{\natexlab{b}})\citenamefont {Cohen}, \citenamefont {Mori-S\'{a}nchez},\ and\ \citenamefont {Yang}}]{Cohen08}%
  \BibitemOpen
  \bibfield  {author} {\bibinfo {author} {\bibfnamefont {A.~J.}\ \bibnamefont {Cohen}}, \bibinfo {author} {\bibfnamefont {P.}~\bibnamefont {Mori-S\'{a}nchez}},\ and\ \bibinfo {author} {\bibfnamefont {W.}~\bibnamefont {Yang}},\ }\href@noop {} {\bibfield  {journal} {\bibinfo  {journal} {Science}\ }\textbf {\bibinfo {volume} {321}},\ \bibinfo {pages} {792} (\bibinfo {year} {2008}{\natexlab{b}})}\BibitemShut {NoStop}%
\bibitem [{\citenamefont {Bajaj}\ \emph {et~al.}(2017)\citenamefont {Bajaj}, \citenamefont {Janet},\ and\ \citenamefont {Kulik}}]{BajajKulik17}%
  \BibitemOpen
  \bibfield  {author} {\bibinfo {author} {\bibfnamefont {A.}~\bibnamefont {Bajaj}}, \bibinfo {author} {\bibfnamefont {J.~P.}\ \bibnamefont {Janet}},\ and\ \bibinfo {author} {\bibfnamefont {H.~J.}\ \bibnamefont {Kulik}},\ }\href@noop {} {\bibfield  {journal} {\bibinfo  {journal} {J. Chem. Phys.}\ }\textbf {\bibinfo {volume} {147}},\ \bibinfo {pages} {191101} (\bibinfo {year} {2017})}\BibitemShut {NoStop}%
\bibitem [{\citenamefont {Wodyński}\ \emph {et~al.}(2021)\citenamefont {Wodyński}, \citenamefont {Arbuznikov},\ and\ \citenamefont {Kaupp}}]{WodynskiArbuzKaupp21}%
  \BibitemOpen
  \bibfield  {author} {\bibinfo {author} {\bibfnamefont {A.}~\bibnamefont {Wodyński}}, \bibinfo {author} {\bibfnamefont {A.~V.}\ \bibnamefont {Arbuznikov}},\ and\ \bibinfo {author} {\bibfnamefont {M.}~\bibnamefont {Kaupp}},\ }\href@noop {} {\bibfield  {journal} {\bibinfo  {journal} {J. Chem. Phys.}\ }\textbf {\bibinfo {volume} {155}},\ \bibinfo {pages} {144101} (\bibinfo {year} {2021})}\BibitemShut {NoStop}%
\bibitem [{\citenamefont {Prokopiou}\ \emph {et~al.}(2022)\citenamefont {Prokopiou}, \citenamefont {Hartstein}, \citenamefont {Govind},\ and\ \citenamefont {Kronik}}]{Prokopiu22}%
  \BibitemOpen
  \bibfield  {author} {\bibinfo {author} {\bibfnamefont {G.}~\bibnamefont {Prokopiou}}, \bibinfo {author} {\bibfnamefont {M.}~\bibnamefont {Hartstein}}, \bibinfo {author} {\bibfnamefont {N.}~\bibnamefont {Govind}},\ and\ \bibinfo {author} {\bibfnamefont {L.}~\bibnamefont {Kronik}},\ }\href@noop {} {\bibfield  {journal} {\bibinfo  {journal} {J. Chem. Theory Comput.}\ }\textbf {\bibinfo {volume} {18}},\ \bibinfo {pages} {2331} (\bibinfo {year} {2022})}\BibitemShut {NoStop}%
\bibitem [{\citenamefont {Bajaj}\ \emph {et~al.}(2022)\citenamefont {Bajaj}, \citenamefont {Duan}, \citenamefont {Nandy}, \citenamefont {Taylor},\ and\ \citenamefont {Kulik}}]{BajajKulik22}%
  \BibitemOpen
  \bibfield  {author} {\bibinfo {author} {\bibfnamefont {A.}~\bibnamefont {Bajaj}}, \bibinfo {author} {\bibfnamefont {C.}~\bibnamefont {Duan}}, \bibinfo {author} {\bibfnamefont {A.}~\bibnamefont {Nandy}}, \bibinfo {author} {\bibfnamefont {M.~G.}\ \bibnamefont {Taylor}},\ and\ \bibinfo {author} {\bibfnamefont {H.~J.}\ \bibnamefont {Kulik}},\ }\href@noop {} {\bibfield  {journal} {\bibinfo  {journal} {J. Chem. Phys.}\ }\textbf {\bibinfo {volume} {156}},\ \bibinfo {pages} {184112} (\bibinfo {year} {2022})}\BibitemShut {NoStop}%
\bibitem [{\citenamefont {Burgess}\ \emph {et~al.}(2023)\citenamefont {Burgess}, \citenamefont {Linscott},\ and\ \citenamefont {O'Regan}}]{BurgessLinscottORegan23}%
  \BibitemOpen
  \bibfield  {author} {\bibinfo {author} {\bibfnamefont {A.~C.}\ \bibnamefont {Burgess}}, \bibinfo {author} {\bibfnamefont {E.}~\bibnamefont {Linscott}},\ and\ \bibinfo {author} {\bibfnamefont {D.~D.}\ \bibnamefont {O'Regan}},\ }\href@noop {} {\bibfield  {journal} {\bibinfo  {journal} {Phys. Rev. B}\ }\textbf {\bibinfo {volume} {107}},\ \bibinfo {pages} {L121115} (\bibinfo {year} {2023})}\BibitemShut {NoStop}%
\bibitem [{\citenamefont {Gál}\ \emph {et~al.}(2009)\citenamefont {Gál}, \citenamefont {Ayers}, \citenamefont {De~Proft},\ and\ \citenamefont {Geerlings}}]{GalAyersProftGeerlings09}%
  \BibitemOpen
  \bibfield  {author} {\bibinfo {author} {\bibfnamefont {T.}~\bibnamefont {Gál}}, \bibinfo {author} {\bibfnamefont {P.~W.}\ \bibnamefont {Ayers}}, \bibinfo {author} {\bibfnamefont {F.}~\bibnamefont {De~Proft}},\ and\ \bibinfo {author} {\bibfnamefont {P.}~\bibnamefont {Geerlings}},\ }\href {https://doi.org/10.1063/1.3233717} {\bibfield  {journal} {\bibinfo  {journal} {The Journal of Chemical Physics}\ }\textbf {\bibinfo {volume} {131}},\ \bibinfo {pages} {154114} (\bibinfo {year} {2009})},\ \Eprint {https://arxiv.org/abs/https://pubs.aip.org/aip/jcp/article-pdf/doi/10.1063/1.3233717/13493895/154114\_1\_online.pdf} {https://pubs.aip.org/aip/jcp/article-pdf/doi/10.1063/1.3233717/13493895/154114\_1\_online.pdf} \BibitemShut {NoStop}%
\bibitem [{Note2()}]{Note2}%
  \BibitemOpen
  \bibinfo {note} {We use here the fact that $\protect \hat {N}$, $\protect \hat {S}^2$, $\protect \hat {S}_z$, and $\protect \hat {H}$ are commuting operators and therefore their eigenstates, $\ket {\Psi _{N,S,M,E}}$, form a complete basis. Among these, only $\ket {\Psi _{N,M}}$ are necessary to describe the ground state $\protect \hat {\Lambda }$, as they have the lowest energy for a given $N$ and $M$. We note in passing that more general ensembles, which include also off-diagonal terms of the form $\ketbra {\Psi _{N_a,M_a}}{\Psi _{N_b,M_b}}$, introduce additional degeneracy to the ground state, but the energy and the spin-densities $n^\sigma _\protect \textrm {ens}(\protect \mathbf {r})$ are unaffected by this extension.}\BibitemShut {Stop}%
\bibitem [{\citenamefont {Lieb}(1983)}]{Lieb}%
  \BibitemOpen
  \bibfield  {author} {\bibinfo {author} {\bibfnamefont {E.~H.}\ \bibnamefont {Lieb}},\ }\href@noop {} {\bibfield  {journal} {\bibinfo  {journal} {Int. J. Quantum Chem.}\ }\textbf {\bibinfo {volume} {24}},\ \bibinfo {pages} {243} (\bibinfo {year} {1983})}\BibitemShut {NoStop}%
\bibitem [{\citenamefont {Janak}(1978)}]{Janak78}%
  \BibitemOpen
  \bibfield  {author} {\bibinfo {author} {\bibfnamefont {J.~F.}\ \bibnamefont {Janak}},\ }\href@noop {} {\bibfield  {journal} {\bibinfo  {journal} {Phys. Rev. B}\ }\textbf {\bibinfo {volume} {18}},\ \bibinfo {pages} {7165} (\bibinfo {year} {1978})}\BibitemShut {NoStop}%
\bibitem [{\citenamefont {Levy}(1995)}]{Levy95}%
  \BibitemOpen
  \bibfield  {author} {\bibinfo {author} {\bibfnamefont {M.}~\bibnamefont {Levy}},\ }\href {https://doi.org/10.1103/PhysRevA.52.R4313} {\bibfield  {journal} {\bibinfo  {journal} {Phys. Rev. A}\ }\textbf {\bibinfo {volume} {52}},\ \bibinfo {pages} {R4313} (\bibinfo {year} {1995})}\BibitemShut {NoStop}%
\bibitem [{Note3()}]{Note3}%
  \BibitemOpen
  \bibinfo {note} {The magnetic field being weak means that $\mu _\protect \textrm {B}B_0$ is smaller than any energy difference in the problem; $f(\protect \mathbf {r})$ is bounded. Therefore, we refer only to terms that are first order in $\protect \mathbf {B}$.}\BibitemShut {Stop}%
\bibitem [{Note4()}]{Note4}%
  \BibitemOpen
  \bibinfo {note} {In our analysis we disregard, for simplicity, the magnetic term $\protect \mathbf {A} \cdot \protect \mathbf {\protect \hat {p}}$, where $\protect \mathbf {A}$ is the electromagnetic vector-potential and $\protect \mathbf {\protect \hat {p}}$ is the momentum operator. [For homogeneous magnetic fields, this term boils down to $\protect \mathbf {B} \cdot \protect \mathbf {\protect \hat {L}}$]. Treatment of this term changes the value of the energy $E(N)$, but it is independent of $M$ -- which is our main focus.}\BibitemShut {Stop}%
\end{thebibliography}%


\begin{thebibliography}{1}

\bibitem{NumPy20}
C.~R. Harris, K.~J. Millman, S.~J. van~der Walt, R.~Gommers, P.~Virtanen, D.~Cournapeau, E.~Wieser, J.~Taylor, S.~Berg, N.~J. Smith, R.~Kern, M.~Picus, S.~Hoyer, M.~H. van Kerkwijk, M.~Brett, A.~Haldane, J.~F. del Río, M.~Wiebe, P.~Peterson, P.~G\'{e}rard-Marchant, K.~Sheppard, T.~Reddy, W.~Weckesser, H.~Abbasi, C.~Gohlke, and T.~E. Oliphant.
\newblock Array programming with numpy.
\newblock {\em Nature}, 585:357, 2020.

\bibitem{Wei}
E.~W. Weisstein.
\newblock {\em CRC Concise Encyclopedia of Mathematics}.
\newblock London: Chapman and Hall, 2003.
\newblock Available on http://mathworld.wolfram.com.

\end{thebibliography}

\end{document}


\title{Supporting information for: \\ \textit{Ensemble ground-state of a many-electron system with fractional electron number and spin: piecewise-linearity and flat plane condition generalized}}

\author{Yuli Goshen}
\affiliation{Fritz Haber Research Center for Molecular Dynamics and Institute of Chemistry, The Hebrew University of Jerusalem, 9091401 Jerusalem, Israel}

\author{Eli Kraisler}
\email[Author to whom correspondence should be addressed: ]{eli.kraisler@mail.huji.ac.il}
\affiliation{Fritz Haber Research Center for Molecular Dynamics and Institute of Chemistry, The Hebrew University of Jerusalem, 9091401 Jerusalem, Israel}

\date{\today}

\maketitle{}

\onecolumngrid  
This document provides further technical details on four subjects. 
In Sec.~\ref{sec:M-dep} we show that certain pure-state expectation values, including the density, $\expval{\hat{n}(\rr)}{\Psi_{N,S,M}}$, do not depend on $M$.
In Sec.~\ref{sec:Higher_moments} we show how the ensemble ground state $\hat{\Gamma}$ and/or the spin distribution $Q_\ens(\rr)$ can be uniquely determined by setting the higher moments of the operator $\hat{S}_z$. In Sec.~\ref{sec:min_DSz} we prove that enforcing the requirement that the deviation $\Delta S_z$ is minimal always fully determines the ground state. 
In Sec.~\ref{sec:Cases_c_d_e} we provide a detailed analysis of Cases (c), (d) and (e), which were introduced in the main text, in absence and in presence of a weak, inhomogeneous magnetic field.

In the following, reference to an equation from the main text starts with the letters MT (e.g., Eq.~(MT:4)), whereas references to equations within this document are by a single number.

\section{Dependence of pure-state quantities on $M$} \label{sec:M-dep}
In this section we clarify which pure-state quantities are independent of the value of $M$, as long as $M \in [-S_\m, S_\m]$, and which are not.  Consider a quantity described by an operator $\hat{P}$ and compare $P(M) = \expval{\hat{P}}{\Psi_{N,M}}$ with, say, $P(M-1)$. Recall that $\hat{S}_\pm \ket{\Psi_{N,M \mp 1}} = \const.  \ket{\Psi_{N,M}}$, where $\hat{S}_\pm=\hat{S}^\dagger_\mp=\hat{S}_x\pm i \hat{S}_y$ are the ladder operators, which raise/lower $M$ by 1. Furthermore, note that $\ket{\Psi_{N,M}}$ is an eigenstate of the operator $\hat{S}_- \hat{S}_+$. Then, $P(M-1) = \expval{\hat{P}}{\Psi_{N,M-1}} = \frac{\expval{\hat{S}_+ \hat{P}\hat{S}_-}{\Psi_{N,M}}}{\expval{\hat{S}_+ \hat{S}_-}{\Psi_{N,M}}}$. It then directly follows that for any operator $\hat{P}$ \emph{that commutes with} $\hat{S}_+$, $P(M-1)=P(M)$, i.e., the value of $P$ is $M$-independent. Such is the case for the Hamiltonian $\hat{H}$, in absence of a magnetic field, for the electron density operator $\hat{n}(\rr)$, for the reduced density matrix, $\hat{\rho}_1(\rr,\rr')$, and for any operator that depends only on spatial, and not on spin-coordinates. However, this is not the case for the Hamiltonian in presence of a magnetic field, for the spin-density operator $\hat{n}_\s(\rr)$ or for the spin-dependent reduced density matrix~$\hat{\gamma}_1(\rr \s,\rr' \s')$.

\section{Higher Moments of $\hat{S}_z$} \label{sec:Higher_moments}

In this section we discuss the possibility to determine the ground state $\hat{\Gamma}$ uniquely by choosing the values of moments of $\hat{S}_z$ (i.e. $\stat{\hat{S}_z},\stat{\hat{S}^2_z},\stat{\hat{S}^3_z}$, etc.), in addition to Eqs.~(\EqOneMT)--(\EqThreeMT), and the possibility to determine the spin distribution 
$Q_\ens(\rr)=n^\up_\ens(\rr)-n^\dw_\ens(\rr)$, by choosing the values of \emph{odd} moments of $\hat{S}_z$. Recall that the first moment is $M_\tot=\stat{\hat{S}_z}$.

First, consider $N_\tot\in\mathbb{N}$ (as relevant in Cases (c) and~(d) in the main text). There are $2S_0+1$ ensemble coefficients, and the moments of~$\hat{S}_z$ are given by 
\begin{equation}\label{eq:moments1}
    \begin{pmatrix}
1\\
M_{\tot}\\
\stat{\hat{S}_{z}^{2}}\\
\vdots\\
\stat{\hat{S}_{z}^{2S_{0}}}
\end{pmatrix}=\begin{pmatrix}1 & \cdots & 1\\
-S_{0} & \cdots & S_{0}\\
\left(-S_{0}\right)^{2} & \cdots & S_{0}^{2}\\
\vdots &  & \vdots\\
\left(-S_{0}\right)^{2S_{0}} & \cdots & S_{0}^{2S_{0}}
\end{pmatrix}\begin{pmatrix}
\gamma_{N_{0},-S_{0}}\\
\gamma_{N_{0},-S_{0}+1}\\
\vdots\\
\vdots\\
\gamma_{N_{0},S_{0}}
\end{pmatrix}.
\end{equation}
The $(2S_0+1)\times(2S_0+1)$ 
matrix of Eq.~\eqref{eq:moments1} is a Vandermonde matrix~\cite{Wei}, where no two columns are identical, hence it is invertible. 
Therefore, the moments of $\hat{S}_z$ up to $\hat{S}_z^{2S_0}$ determine the ensemble coefficients, and thus $\hat{\Gamma}$.

Next we show that $Q_\ens(\rr)$ is determined by the \emph{odd} moments, $\stat{\hat{S}_z^k}$, where $1\leqs k\leqs 2S_0$.
Importantly, the density $n_\ens(\rr)=n^\up_\ens(\rr)+n^\dw_\ens(\rr)$ is unique, depending only on $N_\tot$. Thus, if $Q_\ens(\rr)$ is determined uniquely (while $\hat{\Gamma}$ may remain undetermined), the spin-densities $n^\s_\ens(\rr)$ are determined from $Q_\ens(\rr)$ uniquely, as well.

If for instance $N_\tot$ is an odd integer, then $2S_0$ is an odd integer, and the odd moments of $\hat{S}_z$ up to $\stat{\hat{S}_z^{2S_0}}$ can be expressed as 
\begin{equation}
\label{eq:moments2}
\begin{pmatrix}M_{\tot}\\
\stat{\hat{S}_{z}^{3}}\\
\stat{\hat{S}_{z}^{5}}\\
\vdots\\
\stat{\hat{S}_{z}^{2S_{0}}}
\end{pmatrix}=\begin{pmatrix}\frac{1}{2} & \cdots & S_{0}\\
\left(\frac{1}{2}\right)^{3} & \cdots & S_{0}^{3}\\
\left(\frac{1}{2}\right)^{5} & \cdots & S_{0}^{5}\\
\vdots &  & \vdots\\
\left(-\frac{1}{2}\right)^{2S_{0}} & \cdots & S_{0}^{2S_{0}}
\end{pmatrix}\begin{pmatrix}
\gamma_{N_{0},1/2}-\gamma_{N_{0},-1/2}\\
\gamma_{N_{0},3/2}-\gamma_{N_{0},-3/2}\\
\gamma_{N_{0},5/2}-\gamma_{N_{0},-5/2}\\
\vdots\\
\gamma_{N_{0},S_{0}}-\gamma_{N_{0},-S_{0}}
\end{pmatrix}.
\end{equation} 
The matrix in Eq.~\eqref{eq:moments2} is similar to that of Eq.~\eqref{eq:moments1}, however the powers in each row are not consecutive integers, hence it is not a Vandermonde matrix but a generalized Vandermonde matrix~\cite{Wei}.
This matrix is also 
invertible, meaning that the odd moments $\stat{\hat{S}_z},\ldots,\stat{\hat{S}_z^{2S_0}}$ determine the differences $(\gamma_{N_{0},1/2}-\gamma_{N_{0},-1/2}), \: \ldots, \: (\gamma_{N_{0},S_{0}}-\gamma_{N_{0},-S_{0}})$. 
Since $Q_{N,M}(\rr)=-Q_{N,-M}(\rr)$, the ensemble spin distribution can be expressed as $Q_\ens(\rr)=\sum_{M=\frac{1}{2}}^{S_0} (\gamma_{N_{0},M}-\gamma_{N_{0},-M}) Q_{N,M}(\rr)$, meaning it is also determined by the odd moments $\stat{\hat{S}_z},\ldots,\stat{\hat{S}_z^{2S_0}}$.
One can show in a similar manner that if $N_\tot$ is an even integer then $Q_\ens(\rr)$ is determined by $\stat{\hat{S}_z},\stat{\hat{S}^3_z},\ldots,\stat{\hat{S}_z^{2S_0-1}}$.

Next we consider the case of $N_\tot\notin\mathbb{N}$, where there are now $2(S_0+S_1+1)$ ensemble coefficients.
We express the moments of $\hat{S}_z$, as well as $\alpha$, by

\begin{equation}\label{eq:moments3}
\begin{pmatrix}1\\
\alpha\\
M_{\tot}\\
\stat{\hat{S}_{z}^{2}}\\
\vdots\\
\stat{\hat{S}_{z}^{2S_{0}+2S_{1}}}
\end{pmatrix}=\begin{pmatrix}1 & \cdots & 1 & 1 & \cdots & 1\\
0 & \cdots & 0 & 1 & \cdots & 1\\
-S_{0} & \cdots & S_{0} & -S_{1} & \cdots & S_{1}\\
\left(-S_{0}\right)^{2} & \cdots & S_{0}^{2} & \left(-S_{1}\right)^{2} & \cdots & S_{1}^{2}\\
\vdots &  & \vdots & \vdots &  & \vdots\\
\left(-S_{0}\right)^{2S_{0}+2S_{1}} & \cdots & S_{0}^{2S_{0}+2S_{1}} & -S_{1}^{2S_{0}+2S_{1}} & \cdots & S_{1}^{2S_{0}+2S_{1}}
\end{pmatrix}\begin{pmatrix}\gamma_{N_{0},-S_{0}}\\
\vdots\\
\gamma_{N_{0},S_{0}}\\
\gamma_{N_{0}+1,-S_{1}}\\
\vdots\\
\gamma_{N_{0}+1,S_{1}}
\end{pmatrix}.
\end{equation}

If this matrix is invertible, then indeed the ground state is determined by the values of the moments of $\hat{S}_z$, up to $\stat{\hat{S}_z^{2S_0+2S_1}}$, together with $\alpha$. We have explicitly verified that this is the case for $S_0,S_1\leqs8$, using the \texttt{NumPy}~\cite{NumPy20} library in \texttt{Python}. This covers essentially all the chemically relevant scenarios.
However, if this matrix is not invertible, then it is sufficient to choose the value of $\stat{\hat{S}_{z}^{2S_{0}+2S_{1}+1}}$ as an additional constraint (consistently with all the other constraints). Then,  
\begin{equation}\label{eq:moments4}
\begin{pmatrix}1\\
M_{\tot}\\
\stat{\hat{S}_{z}^{2}}\\
\vdots\\
\stat{\hat{S}_{z}^{2S_{0}+2S_{1}+1}}
\end{pmatrix}=\begin{pmatrix}1 & \cdots & 1 & 1 & \cdots & 1\\
-S_{0} & \cdots & S_{0} & -S_{1} & \cdots & S_{1}\\
\left(-S_{0}\right)^{2} & \cdots & S_{0}^{2} & \left(-S_{1}\right)^{2} & \cdots & S_{1}^{2}\\
\vdots &  & \vdots & \vdots &  & \vdots\\
\left(-S_{0}\right)^{2S_{0}+2S_{1}+1} & \cdots & S_{0}^{2S_{0}+2S_{1}+1} & (-S_{1})^{2S_{0}+2S_{1}+1} & \cdots & S_{1}^{2S_{0}+2S_{1}+1}
\end{pmatrix}\begin{pmatrix}\gamma_{N_{0},-S_{0}}\\
\vdots\\
\gamma_{N_{0},S_{0}}\\
\gamma_{N_{0}+1,-S_{1}}\\
\vdots\\
\gamma_{N_{0}+1,S_{1}}
\end{pmatrix}.
\end{equation}
The matrix of Eq.~\eqref{eq:moments4} is another Vandermonde matrix, hence it is invertible.

To determine $Q_\ens(\rr)$, now with $N_\tot \notin\mathbb{N}$, assume that $N_0$ is odd. Then, analogously to Eq.~\eqref{eq:moments2},
\\
\begin{equation}\label{eq:moments5}
\begin{pmatrix}M_{\tot}\\
\stat{\hat{S}_{z}^{3}}\\
\stat{\hat{S}_{z}^{5}}\\
\vdots\\
\stat{\hat{S}_{z}^{2S_{0}+2S_{1}}}
\end{pmatrix}=\begin{pmatrix}\frac{1}{2} & \cdots & S_{0} & 1 & \cdots & S_{1}\\
\left(\frac{1}{2}\right)^{3} & \cdots & S_{0}^{3} & 1^{3} & \cdots & S_{1}^{3}\\
\left(\frac{1}{2}\right)^{5} & \cdots & S_{0}^{5} & 1^{5} & \cdots & S_{1}^{5}\\
\vdots &  & \vdots & \vdots &  & \vdots\\
\left(\frac{1}{2}\right)^{2S_{0}+2S_{1}} & \cdots & S_{0}^{2S_{0}+2S_{1}} & 1^{2S_{0}+2S_{1}} & \cdots & S_{1}^{2S_{0}+2S_{1}}
\end{pmatrix}\begin{pmatrix}\gamma_{N_{0},{1/2}}-\gamma_{N_{0},-{1/2}}\\
\vdots\\
\gamma_{N_{0},S_{0}}-\gamma_{N_{0},-S_{0}}\\
\gamma_{N_{0}+1,1}-\gamma_{N_{0}+1,-1}\\
\vdots\\
\gamma_{N_{0}+1,S_{1}}-\gamma_{N_{0}+1,-S_{1}}
\end{pmatrix}.
\end{equation}
The matrix in Eq.~\eqref{eq:moments5} is another generalized Vandermonde matrix, meaning it is invertible.
Since
$Q_\ens(\rr)=\sum_{N=N_0}^{N_0+1}\sum_{M>0}(\gamma_{N,M}-\gamma_{N,-M})Q_{N,M}(\rr)$,
we see that $Q_\ens(\rr)$ is fully determined by the odd moments of $\hat{S}_z$, up to~$\stat{\hat{S}_z^{2S_0+2S_1}}$. This can be shown for the case of $N_0$ being even, in the same way. 

\vspace{1cm}

\twocolumngrid

\section{Minimizing $\Delta S_z$}  \label{sec:min_DSz}

In this section we prove that $\hat{\Gamma}$ is determined uniquely by adding the constraint of minimizing $\Delta S_z$, or equivalently, minimizing $\stat{\hat{S}^{2}_z}$, as $\stat{\hat{S}_z} = M_\tot$ is given by Eq.~(\EqTwoMT).
Consider an ensemble $\hat{\Gamma}$ (Eq.~(\EqFiveMT)), which minimizes $\stat{\hat{S}_z^2}$.
Since the ensemble coefficients are all between 0 and 1, we may write them as $\gamma_{N,M}=\sin^2(\theta_{N,M})$. 
Using Lagrange multipliers, with Eqs.~(\EqOneMT)--(\EqThreeMT) as constraints, we find
\begin{widetext}
\begin{align}
    &\frac{\de }{\de \theta_{N,M}}\left(\sum_{N'=N_0}^{N_0+1}\sum_{M'=-S_\m(N')}^{S_\m(N')}\gamma_{N',M'}M'^2\right)=\kappa_{1}\frac{\de }{\de \theta_{N,M}}\left(\sum_{N'=N_0}^{N_0+1}\sum_{M'=-S_\m(N')}^{S_\m(N')}\gamma_{N',M'}\right)+\nonumber\\ &+\kappa_{2}\frac{\de }{\de \theta_{N,M}}\left(\sum_{N'=N_0}^{N_0+1}\sum_{M'=-S_\m(N')}^{S_\m(N')}\gamma_{N',M'}N'\right)+
     \kappa_{3}\frac{\de }{\de \theta_{N,M}}\left(\sum_{N'=N_0}^{N_0+1}\sum_{M'=-S_\m(N')}^{S_\m(N')}\gamma_{N',M'}M'\right)    
\end{align}
\end{widetext}
where $\kappa_1,\kappa_2,\kappa_3$ are some constants. Taking the derivatives, we have $2\sin\theta_{N,M}\cos\theta_{N,M}(M^2-\kappa_1+\kappa_2 N + \kappa_3 M)=0$. Hence, for each $N,M$ such that ${\gamma_{N,M}}\neq0,1$, we have 
\begin{equation}
\label{eq:lagrange}
    M^2=\kappa_1+\kappa_2N+\kappa_3M.
\end{equation}

Since Eq.~\eqref{eq:lagrange} is quadratic in $M$, there are at most two solutions for each value of $N$.
This means there are at most four nonzero coefficients - $\gamma_{N_0,a},\gamma_{N_0,b},\gamma_{N_0+1,c},\gamma_{N_0+1,d}$. Here $M=a,b$ are the solutions to Eq.~\eqref{eq:lagrange} for $N=N_0$, and $M=c,d$ are the solutions for $N=N_0+1$. If Eq.~\eqref{eq:lagrange} has only one solution for $N=N_0$ or $N=N_0+1$, then $a=b$ or $c=d$, respectively.
In addition, from Eq.~\eqref{eq:lagrange} we have 
\begin{equation}
 \label{eq:a+b=c+d}
   a+b=c+d=\kappa_3. 
\end{equation}

\noindent (i) Consider $N_\tot\in\mathbb{N}$. Due to Eqs.~(\EqOneMT),~(\EqThreeMT), we have $\gamma_{N_0+1,c}=\gamma_{N_0+1,d}=0$, and only $\gamma_{N_0,a}$ and $\gamma_{N_0,b}$ may be nonzero. Hence, 
$\hat{\Gamma}=\gamma_{N_0,a}\ketbra{\Psi_{N_0,a}}+\gamma_{N_0,b}\ketbra{\Psi_{N_0,b}}$. 
If $M_\tot$ is an integer (for an even $N_\tot$) or a half-integer (for an odd $N_\tot$), then $\Delta S_z$ is minimized only by $\ketbra{\Psi_{N_\tot,M_\tot}}$, as $\Delta{S}_z=0$ in this case. 
Otherwise, $\gamma_{N_0,a},\gamma_{N_0,b}$ are both nonzero, and $a\neq b$.
From Eqs.~(\EqTwoMT) and~(\EqThreeMT), we have $\gamma_{a}=\frac{1}{a-b}\left(M_{\tot}-b\right)$ and $\gamma_{b}=\frac{1}{a-b}\left(a-M_{\tot}\right)$. 
Therefore \begin{equation}
    \label{eq:S_squared_a_b}
    \stat{\hat{S}_z^2}=\gamma_{N_0,a} a^2+\gamma_{N_0,b} b^2=M_{\tot}\left(a+b\right)-ab.
\end{equation} 
Without loss of generality, we assume that $b<a$. Furthermore, since $M_\tot = \gamma_{N_0,a} a + \gamma_{N_0,b} b$, and all ${\gamma_{N,M}}$'s are within $[0,1]$, we realize that $b<M_\tot<a$. By rewriting Eq.~\eqref{eq:S_squared_a_b} as 
$\stat{\hat{S}_z^2}=\left(M_{\tot}-b\right)a+M_{\tot}b$, we see that $\stat{\hat{S}_z^2}$ is minimized by minimizing $a$. Similarly, by writing 
$\stat{\hat{S}_z^2}=\left(M_{\tot}-a\right)b+M_{\tot}a$, we see that $\stat{\hat{S}_z^2}$ is minimized by maximizing~ $b$. 
Therefore $a$ is the smallest integer or half-integer (in accordance with $N_0$) greater than $M_\tot$, and $b$ is the largest integer/half-integer smaller than $M_\tot$. In particular, $a=b+1$. Therefore, the ground state minimizing $\stat{\hat{S}_z^2}$ is unique in this case.

\noindent (ii) Consider the case $N_\tot\notin\mathbb{N}$. 
We define $\hat{\Gamma}_0=\frac{1}{1-\alpha}(\gamma_{N_0,a}\ketbra{\Psi_{N_0,a}}+\gamma_{N_0,b}\ketbra{\Psi_{N_0,b}})$, $\hat{\Gamma}_1=\frac{1}{\alpha}(\gamma_{N_0+1,c}\ketbra{\Psi_{N_0+1,c}}+\gamma_{N_0+1,d}\ketbra{\Psi_{N_0+1,d}})$, $M_0=\Tr{\hat{\Gamma}_0\hat{S}_z}$ and $M_1=\Tr{\hat{\Gamma}_1\hat{S}_z}$. Then $\hat{\Gamma}=(1-\alpha)\hat{\Gamma}_0+\alpha\hat{\Gamma}_1$, and $\Tr{\hat{\Gamma}\hat{S}_z^2}=(1-\alpha)\Tr{\hat{\Gamma}_0\hat{S}_z^2}+\alpha\Tr{\hat{\Gamma}_1\hat{S}_z^2}$. 
Therefore $a,b$ must minimize $\Tr{\hat{\Gamma}_0\hat{S}_z^2}$ and $c,d$ must minimize  $\Tr{\hat{\Gamma}_1\hat{S}_z^2}$, under the constraints $\Tr{\hat{\Gamma}_0\hat{S}_z}=M_0$ and $\Tr{\hat{\Gamma}_1\hat{S}_z}=M_1$. 

Assume, aiming for contradiction, that 
 $a\neq b$ and $c\neq d$.
Without loss of generality assume that $a>b$ and $c>d$. Therefore, applying the result of (i) to $\hat{\Gamma}_0$ and $\hat{\Gamma}_1$, we find that $a=b+1$ and $c=d+1$, hence $a+b-c-d=2(b-d)$. Since $b$ is the spin of an $N_0$-electron system, and $d$ is the spin of an $N_0+1$-electron system, $b-d$ is a half-integer and therefore $b-d\neq0$, meaning ${a+b-c-d\neq0}$.
This is a contradiction to Eq.~\eqref{eq:a+b=c+d}.
Therefore, by minimizing $\Delta S_z$, we are left with at most three nonzero $\gamma_{N,M}$'s, which can then be derived from Eqs.~(\EqOneMT)--(\EqThreeMT), uniquely. 
Particularly, in case $\gamma_{N_0+1,d}=0$ and is ruled out, Eqs.~(\EqOneMT)--(\EqThreeMT) read:
\begin{equation}
 \begin{pmatrix}
     1\\
     \alpha\\
     M_\tot
 \end{pmatrix}
 =
 \begin{pmatrix}
     1&1&1\\
     0&0&1\\
     a&b&c
 \end{pmatrix}
\begin{pmatrix}
  \gamma_{N_0,a}\\
  \gamma_{N_0,b}\\
  \gamma_{N_0+1,c}
\end{pmatrix}.
\end{equation}
The determinant of the matrix above equals $a-b=1$. Therefore, the matrix is invertible. 
If instead both $\gamma_{N_0+1,c}$ and $\gamma_{N_0+1,d}$ are present, but $\gamma_{N_0,b}$ vanishes, the corresponding matrix is invertible, as well.

Finally, we show that $a$, $b$ and $c$ are uniquely determined. Assume by contradiction that there exists another ensemble state $\hat{\Gamma}'$, which also minimizes $\stat{\hat{S}_z^2}$ under the constraints~(\EqOneMT)--(\EqThreeMT). Thus, there is a  different triad $a'$, $b'$ and $c'$, which corresponds to a minimum. But then we may refer to $\hat{\Gamma}'' = \thalf (\hat{\Gamma}+\hat{\Gamma}')$: it also minimizes $\stat{\hat{S}_z^2}$ under the same constraints, but consists of \emph{more} than three pure states, which is a contradiction. In case the nonzero ensemble coefficients are, e.g., $\gamma_{N_0,a}$, $\gamma_{N_0+1,c}$ and $\gamma_{N_0+1,d}$, the same logic applies.

\twocolumngrid

\section{Analysis of Cases (c), (d) and (e)} \label{sec:Cases_c_d_e}

In this section we present in detail Cases (c), (d) and (e), which were introduced in the main text. We start by considering spin migration at integer electron number, namely, the situation when $N_\tot = N_0 = \const.$, i.e. $\alpha = 0$ and $M_\tot \in [-S_0, S_0]$. As mentioned in the main text, from Eq.~(\EqNineMT) we realize that $\sum_{M=-S_1}^{S_1} \gamma_{N_0+1,M} = 0$, and since $\gamma_{N,M} \in [0,1]$, it follows that $\gamma_{N_0+1,M} = 0$ for all $M$. As we are left with $\gamma_{N_0,M}$ only, for brevity, we may omit in what follows the index~$N_0$. Yet, when it serves clarity we add it back. From Eq.~(\EqNineMT) we then get $\sum_{M=-S_0}^{S_0} \gamma_M = 1$ and $\sum_{M=-S_0}^{S_0} M \gamma_M = M_\tot$ -- two equations to determine $(2S_0+1)$ coefficients. We therefore expect that for $S_0 \geqs 1$ the coefficients $\gamma_{N,M}$, and therefore the ensemble state $\hat{\Gamma}$, would not be determined unambiguously.

We now refer to specific scenarios of spin migration, for selected values of $S_0$. Trivially, for $S_0 = 0$, there is no spin migration at all, $\gamma_0 = 1$ and $M_\tot$ must equal 0.

\vspace{2ex}
\noindent \textbf{For} $\boldsymbol{S_0 = \half}$ (e.g., spin migration for the H or for the Li atom), which is a particular scenario within Case (a),  we find that
\begin{align}
    \gamma_{1/2} = & \half + M_\tot \nonumber \\
    \gamma_{-1/2} = & \half - M_\tot. 
\end{align}
The coefficients are determined unambiguously, and so are all the quantities that describe the system. In particular, $E_\ens(N_\tot,M_\tot) = E(N_0)$, $\stat{\hat{S}_z} = M_\tot$,  $\stat{\hat{S}_z^2} = \tfrac{1}{4}$ -- independent of $M_\tot$. More generally, all the even moments of $\hat{S}_z$, $\stat{\hat{S}_z^{2k}} = (\thalf)^{2k}$ are $M_\tot$-independent and all the odd moments $\stat{\hat{S}_z^{2k+1}} = (\thalf)^{2k} M_\tot$ are linear in $M_\tot$ ($k \in \mathbb{N}$). The deviation in $\hat{S}_z$ is $\Delta S_z \defeq \sqrt{\stat{\hat{S}_z^2} - \stat{\hat{S}_z}^2} = \sqrt{\tfrac{1}{4} - M_\tot^2}$.  

To find the ensemble spin-densities, $n^\s_\ens(\rr)$, we recall that: (i) for any pure-state, $\ket{\Psi_{N,M}}$, the spin-densities relate as $n^\dw_{N,M}(\rr) = n^\up_{N,-M}(\rr)$; (ii) the total density is $M$-independent (see Sec.~\ref{sec:M-dep} above). Then,
\begin{equation} \label{eq:n_s__S0_1_2}
    n^\s_\ens(\rr) = \half n_{N_0}(\rr) + \half \delta_\s \frac{M_\tot}{S_0} Q_{N_0,S_0}(\rr),
\end{equation}
where 
\begin{equation}
\delta_\s=  \lp\{         \begin{array}{rcl}
                1      & : & \s = \up \\
                -1 & : & \s = \dw
              \end{array}
           \right. ,   
\end{equation}
and $Q(\rr) \defeq n^\up(\rr) - n^\dw(\rr)$ is the spin distribution. From Eq.~\eqref{eq:n_s__S0_1_2} it directly follows that the total ensemble density $n_\ens(\rr) = n_{N_0}(\rr)$ is $M_\tot$-independent and the spin distribution
\begin{equation} \label{eq:Q_ens_migration}
    Q_\ens(\rr) = \frac{M_\tot}{S_0} Q_{N_0,S_0}(\rr),
\end{equation}
is linear in $M_\tot$. For all cases of integer $N_\tot$, the spin polarization, $\zeta(\rr) = \frac{n^\up(\rr) - n^\dw(\rr)}{n^\up(\rr) + n^\dw(\rr)}$, has an analogous form to that of $Q_\ens(\rr)$.

\vspace{2ex}
\noindent \textbf{For} $\boldsymbol{S_0 = 1}$ (e.g., spin migration for the C atom), which is \textbf{Case (c)} of the main text, we obtain three coefficients that are not identically zero. They can be expressed as:
\begin{align}
    \gamma_1   = & \half x +\half M_\tot \nonumber \\
    \gamma_0   = & 1-x                             \\
    \gamma_{-1}= & \half x -\half M_\tot \nonumber
\end{align}
where $x$ remains undetermined. The value of $x$ is not arbitrary, though: it is confined by the requirement that all $\gamma_{N,M} \in [0,1]$. Applying this requirement to all three coefficients above, we find that $x \in [ \, |M_\tot|,1]$.

Whereas the total energy, $E_\ens(N_\tot,M_\tot) = E(N_0)$ and all the odd moments of $\hat{S}_z$, $\stat{\hat{S}_z^{2k+1}} = M_\tot$, are determined uniquely, the even moments of $\hat{S}_z$ equal $\stat{\hat{S}_z^{2k}} = x$. Therefore, to fully determine the ensemble ground state for this Case, it is possible to add a requirement as to the value of, e.g., $\stat{\hat{S}_z^2}$ (in agreement with the conclusions of Sec.~\ref{sec:Higher_moments} above). 

\begin{figure}
    \centering
    \includegraphics[width=0.4\textwidth]{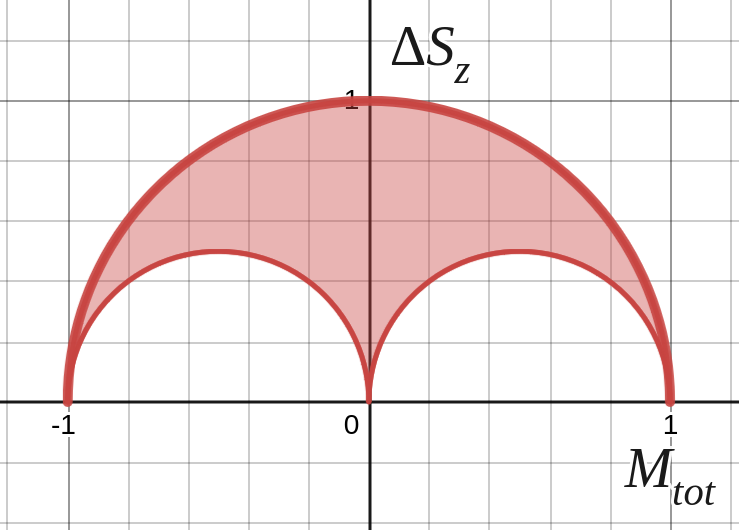}
    \caption{The deviation $\Delta S_z$ versus $M_\tot$, for $S_0=1$. Minimal deviation $(\Delta S_z)_\m$: solid thin line, maximal deviation $(\Delta S_z)_{\max}$: solid thick line.}
    \label{fig:DSz_S_1}
\end{figure}

\begin{figure}
    \centering
    \includegraphics[width=0.4\textwidth]{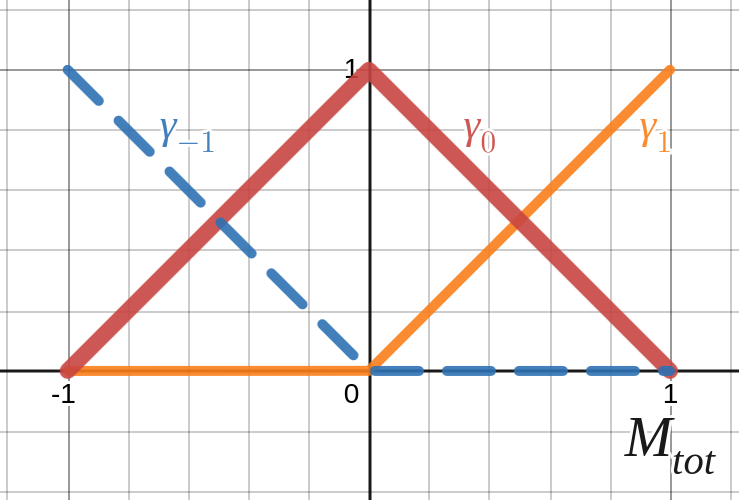}
    \caption{The ensemble coefficients, $\gamma_{-1}$ (dashed blue), $\gamma_{0}$ (solid thick red) and $\gamma_{1}$ (solid orange), versus $M_\tot$, for $S_0=1$, for the scenario of $(\Delta S_z)_\m$.}
    \label{fig:gammas_S_1}
\end{figure}

Alternatively, it is possible to restrict the deviation, $\Delta S_z$, to be minimal. Accounting for the aforementioned range of the variable $x$, we find that $(\Delta S_z)_\m = \sqrt{|M_\tot| - M_\tot^2}$, as depicted in Fig.~\ref{fig:DSz_S_1}.
The ensemble coefficients in this case equal 
\begin{align}
    \gamma_1   & =  \half (|M_\tot| + M_\tot) =  \lp\{ 
        \begin{array}{rcl}
                0      & : & M_\tot < 0 \\
                M_\tot & : & M_\tot \geqs 0
              \end{array}
           \right.  \nonumber \\
    \gamma_0   & =  1- |M_\tot|  \\
    \gamma_{-1} &=  \half (|M_\tot| - M_\tot) =  \lp\{         
        \begin{array}{rcl}
                -M_\tot & : & M_\tot < 0 \\
                0 & : & M_\tot \geqs 0
                \end{array}
           \right. \nonumber 
\end{align}
(see Fig.~\ref{fig:gammas_S_1}).
Note that for every value of $M_\tot$ we obtain two nonzero ensemble coefficients at most. Therefore, as we increase $M_\tot$ from -1 to 0, we see a linear transition from the pure state $\ketbra{\Psi_{N_0,-1}}$ to $\ketbra{\Psi_{N_0,0}}$; further increase of $M_\tot$ from 0 to 1 gives a linear transition from $\ketbra{\Psi_{N_0,0}}$ to $\ketbra{\Psi_{N_0,1}}$. At the points $M_\tot = -1, \: 0, \: 1$, we have pure states (only one coefficient is nonzero) and therefore the deviation $\Delta S_z$ vanishes (see Fig.~\ref{fig:DSz_S_1}).

For completeness, consider the scenario where $\Delta S_z$ is taken to be maximal. Then, $x=1$ and $(\Delta S_z)_{\max} = \sqrt{1 - M_\tot^2}$ (depicted in Fig.~\ref{fig:DSz_S_1}, as well). For such a case, $\gamma_1 = \thalf(1+M_\tot)$, $\gamma_0 = 0$, $\gamma_{-1} = \thalf(1-M_\tot)$, and we obtain, upon increase of $M_\tot$, a gradual transition between $\ketbra{\Psi_{N_0,-1}}$ and $\ketbra{\Psi_{N_0,1}}$ (excluding $\ketbra{\Psi_{N_0,0}}$).

The spin-densities for this Case can be written at first as $ n^\s_\ens(\rr) = \half (1-x) n_{N_0,0}(\rr) + \half x \, n_{N_0,1}(\rr) + \half \delta_\s M_\tot Q_{N_0,1}(\rr)$, but since the pure-state densities are $M$-independent, it directly follows that, similarly to the $S_0=\thalf$ case, $n^\s_\ens(\rr)$ satisfies Eq.~\eqref{eq:n_s__S0_1_2} and does not depend on $x$,
and $Q_\ens(\rr)$ satisfies Eq.~\eqref{eq:Q_ens_migration}. 

In presence of a magnetic field, the total energy in Case~(c) changes linearly with $M_\tot$, as described in the main text. It does not depend, however, on $x$, and therefore cannot be minimized with respect to $x$. Surprisingly, introduction of a magnetic field does not remove the ambiguity in the ground state, for this Case. Therefore, in this respect, the requirement on $\Delta S_z$ is stronger.

\vspace{2ex}
\noindent \textbf{For} $\boldsymbol{S_0 = \tfrac{3}{2}}$ (e.g., spin migration for the N atom), which is \textbf{Case (d)} of the main text, we obtain four coefficients, which are not identically zero. They can be expressed using the variables $x$ and $y$:
\begin{align}
    \gamma_{3/2}  = & \half x +\frac{1}{3} (M_\tot-y) \nonumber \\
    \gamma_{1/2}  = & \half(1-x) + y                             \\
    \gamma_{-1/2} = & \half(1-x) - y                   \nonumber \\
    \gamma_{-3/2} = & \half x -\frac{1}{3} (M_\tot-y). \nonumber 
\end{align}
Whereas $E_\ens(N_\tot,M_\tot) = E(N_0)$ and $\stat{\hat{S}_z} = M_\tot$ are determined uniquely, other quantities that describe the system are not. The even moments of $\hat{S}_z$ equal $\stat{\hat{S}_z^{2k}} = S_0^{2k} x + (S_0-1)^{2k} (1-x)$, and are linear in $x$ and independent of $y$, whereas the odd moments of $\hat{S}_z$ are $\stat{\hat{S}_z^{2k+1}} = S_0^{2k} (M_\tot-y) + (S_0-1)^{2k} y$ and are linear in $y$ and independent of $x$. To fully determine the ensemble ground state for this Case, we can add two constraints, e.g., for $\stat{\hat{S}_z^2}$ and $\stat{\hat{S}_z^3}$. 

As opposed to previous cases, here even the spin-densities are not determined unambiguously. They can be expressed as 
\begin{align} \label{eq:n_s_ens__d}
n^\s_\ens(\rr) &= \half n_{N_0}(\rr) + \half \delta_\s M_\tot \frac{Q_{N_0,3/2}(\rr)}{3/2} + \nonumber \\
&+ \half \delta_\s y \lp( \frac{Q_{N_0,1/2}(\rr)}{1/2} - \frac{Q_{N_0,3/2}(\rr)}{3/2}\rp),
\end{align}
being independent of $x$ and linear in $y$. The spin distribution $Q_\ens(\rr)$ can then be written as
\begin{align}
Q_\ens(\rr) =  (M_\tot-y) \frac{Q_{N_0,3/2}(\rr)}{3/2} + y \frac{Q_{N_0,1/2}(\rr)}{1/2};
\end{align}
it is linear in $y$, as well. For $y=0$ it reduces to Eq.~\eqref{eq:Q_ens_migration}.

\begin{figure*}
    \centering
    \includegraphics[width=0.1575\textwidth]{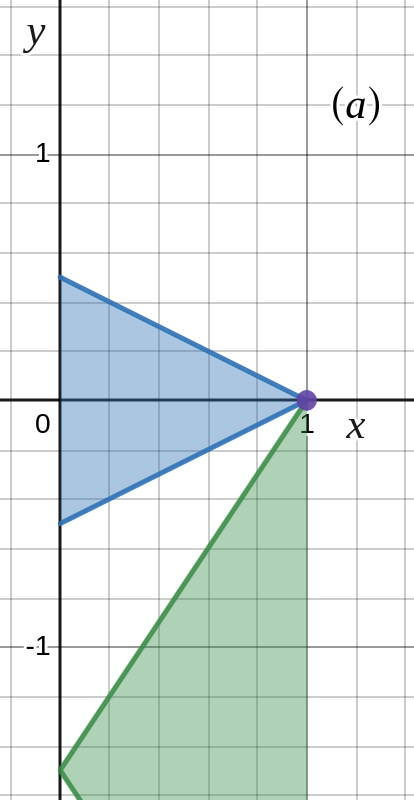}
    \includegraphics[width=0.16\textwidth]{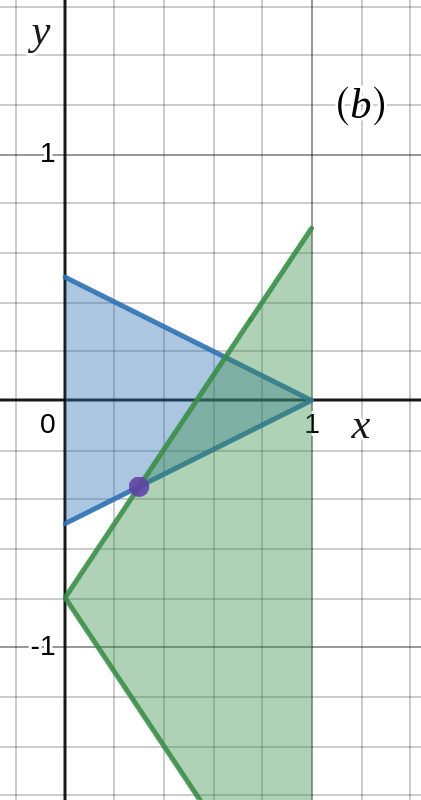}
    \includegraphics[width=0.1585\textwidth]{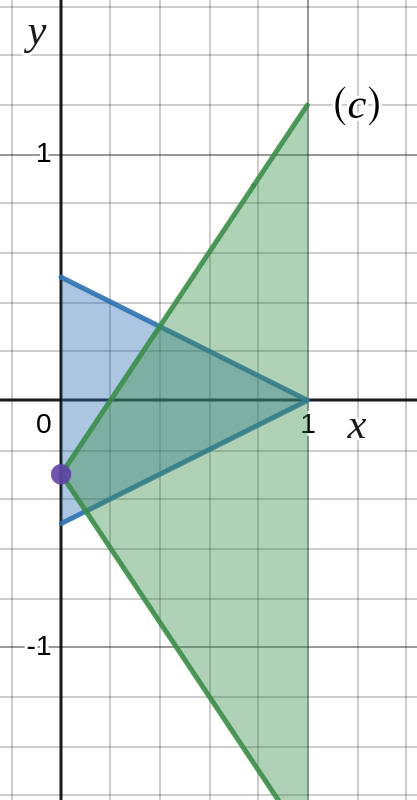}
    \includegraphics[width=0.157\textwidth]{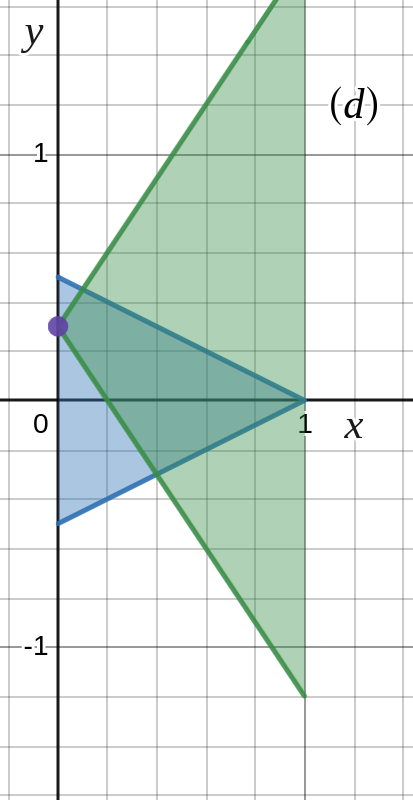}
    \includegraphics[width=0.158\textwidth]{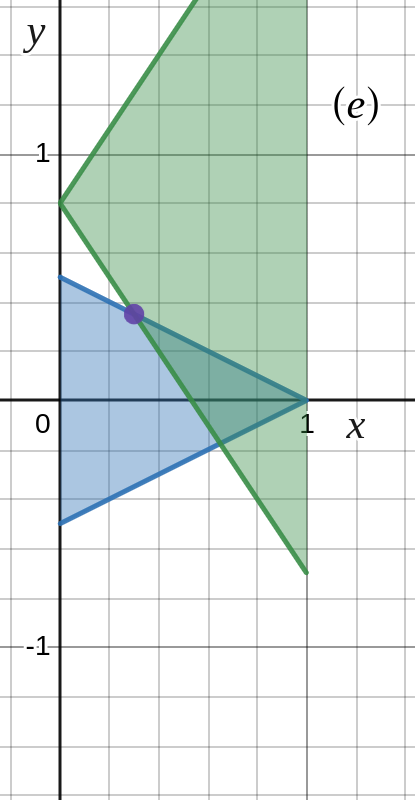}
    \includegraphics[width=0.161\textwidth]{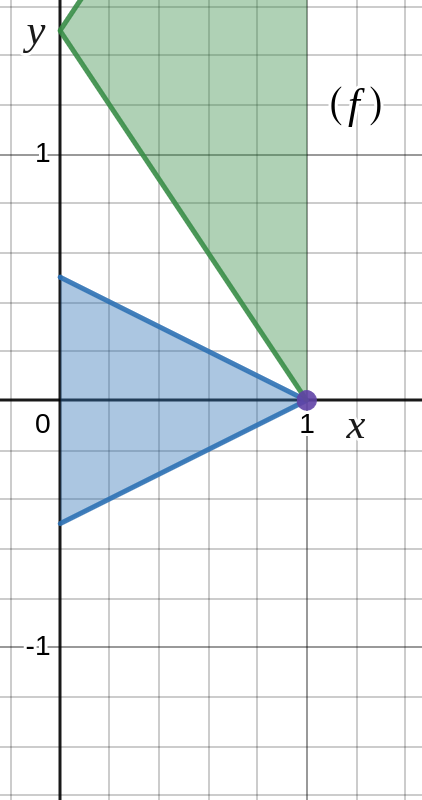}
    \caption{The allowed domain of $x$ and $y$, for $S_0=\tfrac{3}{2}$, is the intersection of the green and blue areas. In (a), $M_\tot = -1.5$; in (b), $M_\tot = -0.8$; in (c), $M_\tot = -0.3$; in (d), $M_\tot = 0.3$; in (e), $M_\tot = 0.8$; in (f), $M_\tot = 1.5$. The lowest possible value of $x$ is denoted by a purple point. For an interactive plot, see: \url{https://www.desmos.com/calculator/bndzdulimf}. }
    \label{fig:x-y_domain_S=3/2}
\end{figure*}

As we already mentioned above, one of the ways to fully determine the ground state 
is to require that the deviation $\Delta S_z$ is minimal. In our Case,  $\Delta S_z = \sqrt{\tfrac{1}{4} - M_\tot^2 + 2x}$, which means that the minimal value of $\Delta S_z$ is obtained for the minimal value of $x$.  As before, the domain of $x$ (and $y$) is restricted by the requirement that all the ensemble coefficients are within $[0,1]$. Superimposing all eight conditions that emerge from this requirement, we are left with two inequalities: $|y| \leqs \thalf (1-x)$ and $|y-M_\tot| \leqs \tfrac{3}{2} x$. Graphically, the domain of $x$ and $y$ (which parametrically depends on $M_\tot$) is illustrated in Fig.~\ref{fig:x-y_domain_S=3/2}. 

\begin{figure}
    \centering
    \includegraphics[width=0.37\textwidth]{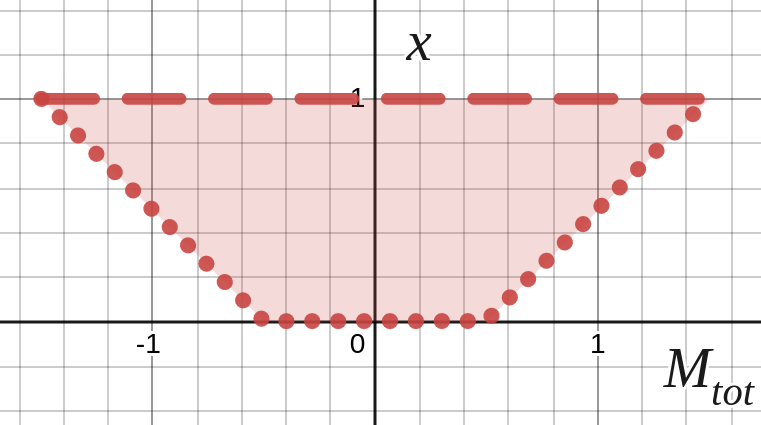}
    \caption{Dependence of the minimal and the maximal values of $x$ ($x_\m$, dotted and $x_{\max}$, dashed) on $M_\tot$, for $S_0 = \tfrac{3}{2}$. }
    \label{fig:x_min_max__S_1_5}
\end{figure}

\begin{figure}
    \centering
    \includegraphics[width=0.4\textwidth]{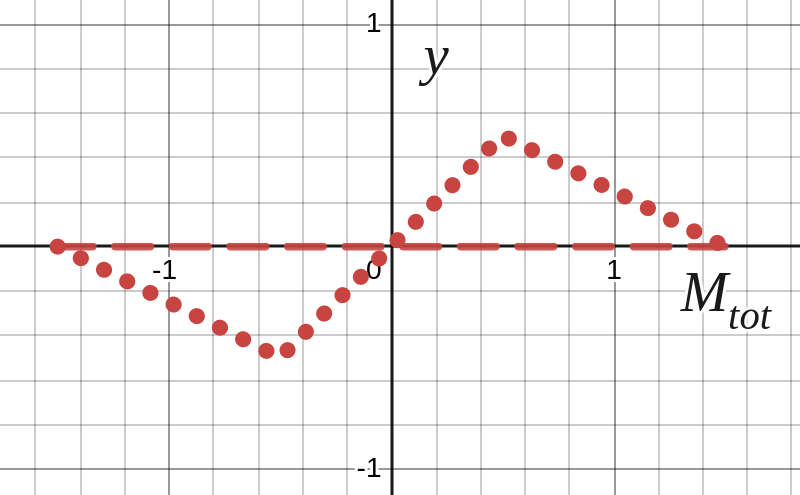}
    \caption{Dependence of the $y$-values, which correspond to the minimal and the maximal values of $x$ ($y_0=y(x_\m)$, dotted and $y_1=y(x_{\max})$, dashed) on $M_\tot$, for $S_0 = \tfrac{3}{2}$. }
    \label{fig:y_1_2__S_1_5}
\end{figure}

From Fig.~\ref{fig:x-y_domain_S=3/2} we conclude that the lowest value of $x$ is
\begin{align}
    x_\m(M_\tot) = \lp\{ 
        \begin{array}{rcl}
                -M_\tot - \half & : & M_\tot \in [-\frac{3}{2}, -\half]\\[3pt]
                0               & : & M_\tot \in [-\half, \half]\\[3pt]
                M_\tot - \half  & : & M_\tot \in [\half, \frac{3}{2}]
              \end{array}.
           \right.  \nonumber \\
\end{align}
Observing the geometry of the $x-y$ domain, we find that for each value of $x_\m$ there is \emph{one} corresponding value of $y$:
\begin{align}
    y_0(M_\tot) \defeq y(x_\m) = \lp\{ 
        \begin{array}{rcl}
                -\frac{3}{4} - \half M_\tot & : & M_\tot \in [\shm \frac{3}{2}, \shm \half]\\[3pt]
                M_\tot                      & : & M_\tot \in [\shm \half, \half]\\[3pt]
                \frac{3}{4} - \half M_\tot  & : & M_\tot \in [\half, \frac{3}{2}]
              \end{array}.
           \right.  \nonumber \\
\end{align}
Therefore, by minimization of $\Delta S_z$, the ground state is defined completely, in agreement with Sec.~\ref{sec:min_DSz}.
The dependence of $x_\m$ and $y_0$ on $M_\tot$ is depicted on Figs.~\ref{fig:x_min_max__S_1_5} and~\ref{fig:y_1_2__S_1_5}, respectively.

For completeness, let us note that the maximal value of $x$ is $x_{\max}=1$, for any $M_\tot$, and $y_1=y(x_{\max})=0$. 
Therefore, in this case also maximizing $\Delta S_z$ sets both $x$ and $y$ and fully determines the ensemble ground state. 

\begin{figure}
    \centering
    \includegraphics[width=0.4\textwidth]{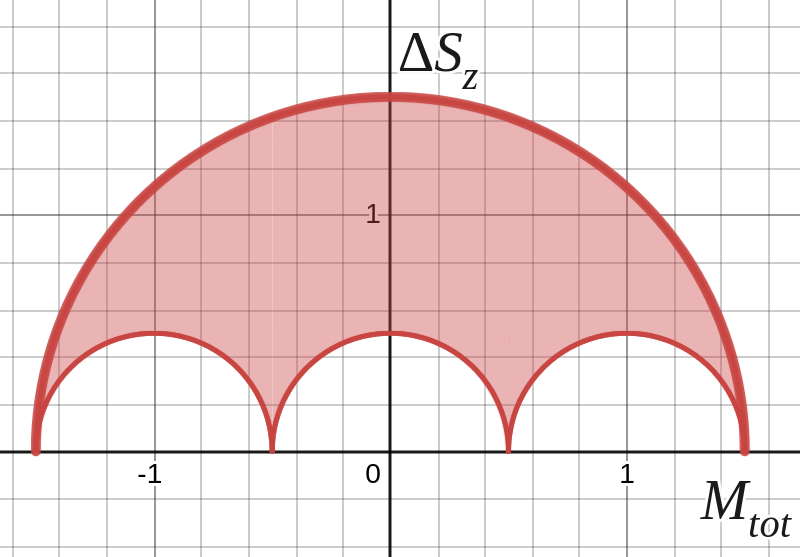}
    \caption{Deviation $\Delta S_z$ versus $M_\tot$, for $S_0=\tfrac{3}{2}$. Minimal deviation, $(\Delta S_z)_\m$: solid thin line, maximal deviation $(\Delta S_z)_{\max}$: solid thick line.}
    \label{fig:DSz_min_max_S_1_5}
\end{figure}

The minimal value of $\Delta S_z$ is 
\begin{align}
    (\Delta S_z)_\m = \lp\{ 
        \begin{array}{lcl}
                \sqrt{(M_\tot+\thalf)(-M_\tot-\tfrac{3}{2})} \!\!\! & : & \!\! M_\tot \! \in \! [\shm \frac{3}{2}, \shm \half]\\
                \sqrt{\tfrac{1}{4} - M_\tot^2}  \!\!\! & : & \!\! M_\tot \! \in \! [\shm \half, \half]\\
                \sqrt{(M_\tot-\thalf)(-M_\tot+\tfrac{3}{2})}  \!\!\! & : & \!\! M_\tot \! \in \!  [\half, \frac{3}{2}]
              \end{array}
           \right.  \nonumber \\
\end{align}
and the maximal value is $(\Delta S_z)_{\max{}} = \sqrt{\tfrac{9}{4} - M_\tot^2}$; see Fig.~\ref{fig:DSz_min_max_S_1_5}.

\begin{figure}
    \centering
    \includegraphics[width=0.5\textwidth]{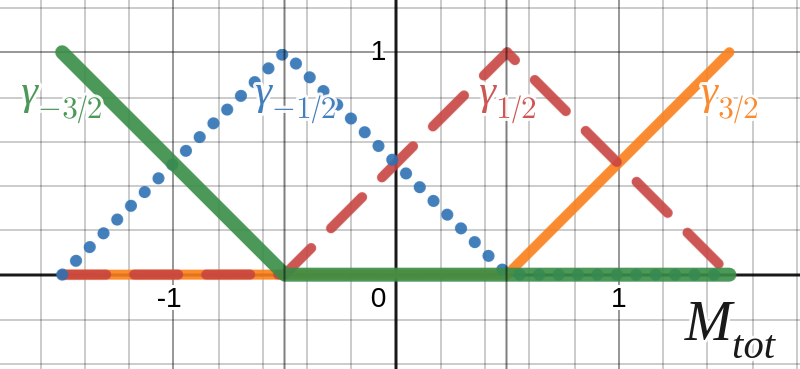}
    \caption{Ensemble coefficients, $\gamma_{-3/2}$ (thick solid green), $\gamma_{-1/2}$ (dotted blue), $\gamma_{1/2}$ (dashed red) and $\gamma_{3/2}$ (solid orange), versus $M_\tot$, for $S_0=\tfrac{3}{2}$, for the scenario of $(\Delta S_z)_\m$.}
    \label{fig:gammas_S_1_5}
\end{figure}

For the scenario of minimal $\Delta S_z$, the ensemble coefficients equal
\begin{align}
\gamma_{3/2}   & = \lp\{ 
    \begin{array}{lcl}
        0             & : & M_\tot \in [-\frac{3}{2},-\half]\\[3pt]
        0             & : & M_\tot \in [-\half, \half]\\[3pt]
        M_\tot-\half  & : & M_\tot \in [\half, \frac{3}{2}]
    \end{array}
                \right.  \nonumber \\
\gamma_{1/2}   & = \lp\{ 
    \begin{array}{lcl}
        0             & : & M_\tot \in [-\frac{3}{2},-\half]\\[3pt]
        \half+M_\tot      & : & M_\tot \in [-\half, \half]\\[3pt]
        \frac{3}{2}-M_\tot & : & M_\tot \in [\half, \frac{3}{2}]
    \end{array}
                \right.  \\
\gamma_{-1/2}   & = \lp\{ 
    \begin{array}{lcl}
        \frac{3}{2}+M_\tot & : & M_\tot \in [-\frac{3}{2},-\half]\\[3pt]
        \half-M_\tot           & : & M_\tot \in [-\half, \half]\\[3pt]
        0             & : & M_\tot \in [\half, \frac{3}{2}]
    \end{array}
                \right.  \nonumber \\
\gamma_{-3/2}   & = \lp\{ 
    \begin{array}{lcl}
        -\half - M_\tot & : & M_\tot \in [-\frac{3}{2},-\half]\\[3pt]
        0             & : & M_\tot \in [-\half, \half]\\[3pt]
        0             & : & M_\tot \in [\half, \frac{3}{2}]
    \end{array}
                \right.  \nonumber 
\end{align}
and are depicted in Fig.~\ref{fig:gammas_S_1_5}.
Similarly to Case (c), for every value of $M_\tot$ we obtain two nonzero ensemble coefficients, at most. Therefore, as we increase $M_\tot$, %
we find a linear transition from $\ketbra{\Psi_{N_0,-3/2}}$ to $\ketbra{\Psi_{N_0,-1/2}}$ and then from $\ketbra{\Psi_{N_0,-1/2}}$ to $\ketbra{\Psi_{N_0,1/2}}$ %
and finally from $\ketbra{\Psi_{N_0,1/2}}$ to $\ketbra{\Psi_{N_0,3/2}}$. %

The behavior for the moments of $\hat{S}_z$ is %
piecewise-linear in $M_\tot$. The even moments are
\begin{align}
&\stat{\hat{S}_z^{2k}}  = \nonumber\\
&\lp\{ 
    \begin{array}{lcl}
        (S_0 \! \shm \! 1)^{2k}(\tfrac{3}{2} \! + \! M_\tot) \shm S_0^{2k}(\thalf \! + \! M_\tot) \!\!\! & : & M_\tot \in [\shm \frac{3}{2}, \shm 
        \half]\\[3pt]
        (S_0 \! \shm \! 1)^{2k}  & : & M_\tot \in [\shm \half, \half]\\[3pt]
        (S_0 \! \shm \! 1)^{2k}(\tfrac{3}{2} \! \shm \! M_\tot) +S_0^{2k}(M_\tot \shm \thalf)  & : & M_\tot \in [\half, \frac{3}{2}]
    \end{array}
\right. 
\end{align}
and have the muffin-tin shape, whereas the odd moments are
\begin{align}
&\stat{\hat{S}_z^{2k+1}}  = \nonumber\\
&\!\!\lp\{ 
    \begin{array}{lcl}
         \!\!\! \shm (S_0 \!\shm\! 1)^{2k+1} %
         \!\shp\!\! \lp[ S_0^{2k+1} \!\!\shm\! (S_0 \!\shm\! 1)^{2k+1} \rp] \!\! (M_\tot \shp \half) \!\!\! & : & \!\!\! M_\tot \!\!\in\!\! [\shm \frac{3}{2}, \!\!\shm \half]\\[5pt]
        (S_0 \shm 1)^{2k} M_\tot \!\!\! & : & \!\!\! M_\tot \!\!\in\!\! [\shm \half, \half]\\[5pt]
        (S_0 \!\shm\! 1)^{2k+1} %
        \!\!+\!\!\lp[ S_0^{2k+1} \!\!\shm\! (S_0 \!\shm\! 1)^{2k+1} \rp] \!\! (M_\tot \!\!\shm\! \half)  \!\!\! & : & \!\!\! M_\tot \!\!\in\!\! [\half, \frac{3}{2}]
    \end{array}.
\right. 
\end{align}
Finally, the spin distribution for the case of minimal $\Delta S_z$ equals
\begin{align}
&Q_\ens(\rr)  = \nonumber\\
&\!\!\lp\{ 
    \begin{array}{lcl}
         \!\shm Q_{1/2}(\rr) + (M_\tot+\thalf) (Q_{3/2}(\rr) \!\shm\! Q_{1/2}(\rr)) \!\!\! & : & \!\!\! M_\tot \!\!\in\!\! [\shm\frac{3}{2}, \!\shm \half]\\[3pt]
         \frac{M_\tot}{1/2} Q_{1/2}(\rr) \!\!\! & : & \!\!\! M_\tot \!\!\in\!\! [\shm \half, \half]\\[3pt]
         Q_{1/2}(\rr) + (M_\tot \shm \thalf) (Q_{3/2}(\rr) \shm  Q_{1/2}(\rr)) \!\!\! & : & \!\!\! M_\tot \!\!\in\!\! [\half, \frac{3}{2}]
    \end{array}.
\right. 
\end{align}
Interestingly, $Q_\ens(\rr)$, and therefore also $\zeta_\ens(\rr)$, are \emph{piecewise-linear} in $M_\tot$, for this scenario.

For completeness, consider the case of maximal $\Delta S_z$, which is obtained, as we mentioned above, for $x=1$ and $y=0$. Then, the ensemble coefficients are $\gamma_{3/2} = \thalf + \tfrac{1}{3} M_\tot$, $\gamma_{1/2} = \gamma_{-1/2} = 0$ and $\gamma_{-3/2} = \thalf - \tfrac{1}{3} M_\tot$. Therefore, as we increase $M_\tot$, we obtain a linear transition between $\ketbra{\Psi_{N_0,-3/2}}$ and $\ketbra{\Psi_{N_0,3/2}}$. The even moments of $\hat{S}_z$ equal $\stat{\hat{S}_z^{2k}} = S_0^{2k}$ and are constant, whereas the odd moments $\stat{\hat{S}_z^{2k+1}} = S_0^{2k} M_\tot$ are linear in $M_\tot$.  %
The spin-densities $n^\s_\ens(\rr)$ and the spin distribution $Q_\ens(\rr)$ satisfy Eqs.~\eqref{eq:n_s__S0_1_2} and~\eqref{eq:Q_ens_migration}, respectively, with $S_0 = \tfrac{3}{2}$.

In presence of a magnetic field $\mathbf{B}(\rr) = B_0 f(\rr) \hat{z}$ in Case (d), the total energy becomes $\tilde{E}_\ens(N_0,M_\tot) = E(N_0) +  \muB B_0 M_\tot F_{N_0,3/2} +  \muB B_0 (F_{N_0,1/2}-F_{N_0,3/2}) y$. It can be minimized then with respect to $y$. As explained in the main text, we are looking then for the lowest and the highest possible values of $y$ (see Fig.~\ref{fig:x-y_domain_S=3/2}). 

\begin{figure}
    \centering
    \includegraphics[width=0.4\textwidth]{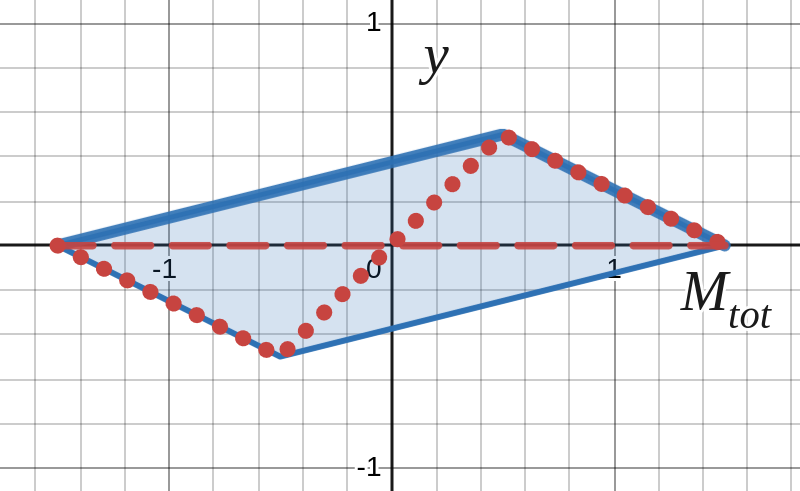}
    \caption{Dependence of the minimal and the maximal values of $y$ ($y_\m$, solid blue and $y_{\max}$, solid thick blue) on $M_\tot$, for $S_0 = \tfrac{3}{2}$. The functions $y_0(M_\tot)=y(x_\m(M_\tot))$ (dotted red) and $y_1(M_\tot)=y(x_{\max}(M_\tot))$ (dashed red) are shown for comparison. }
    \label{fig:y_min_max_0_1}
\end{figure}

\begin{figure}
    \centering
    \includegraphics[width=0.4\textwidth]{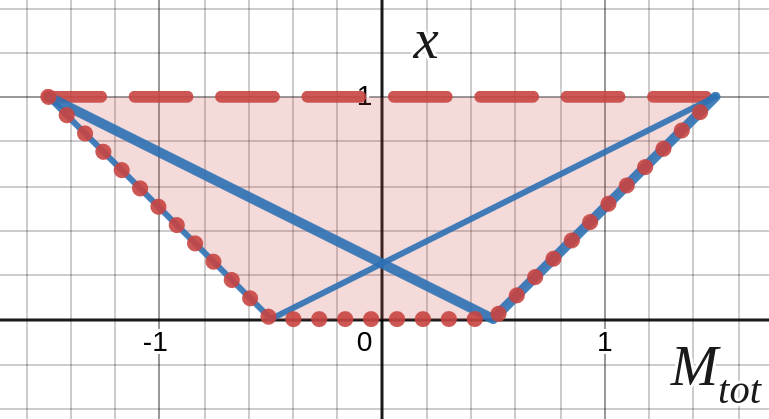}
    \caption{Dependence of the $x$-values $x_1=x(y_\m)$ (solid blue) and $x_2=x(y_{\max})$ (solid thick blue) on $M_\tot$, for $S_0 = \tfrac{3}{2}$. The functions $x_\m(M_\tot)$ (dotted red) and $x_{\max}(M_\tot)$ (dashed red) are shown for comparison. }
    \label{fig:x_min_max_1_2_S_1_5}
\end{figure}

\begin{figure}
    \centering
    \includegraphics[width=0.45\textwidth]{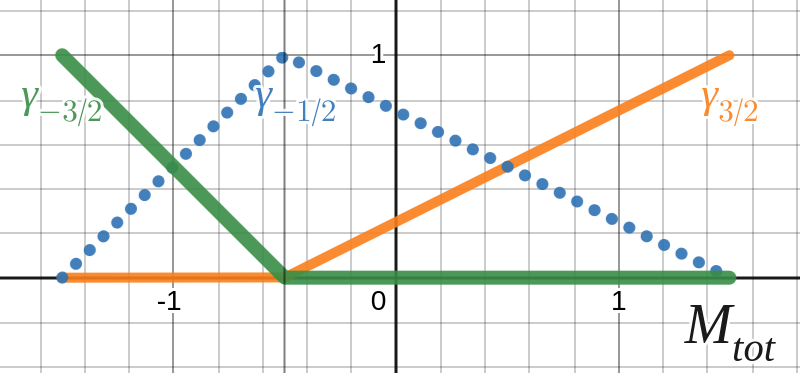}
    \caption{Ensemble coefficients, $\gamma_{-3/2}$ (thick solid green), $\gamma_{-1/2}$ (dotted blue), and $\gamma_{3/2}$ (solid orange), versus $M_\tot$, for $S_0=\tfrac{3}{2}$, for the scenario of $y_\m$. $\gamma_{1/2}=0$ is not shown. }
    \label{fig:gammas_S_3_2_y_min}
\end{figure}

\begin{figure}
    \centering
    \includegraphics[width=0.45\textwidth]{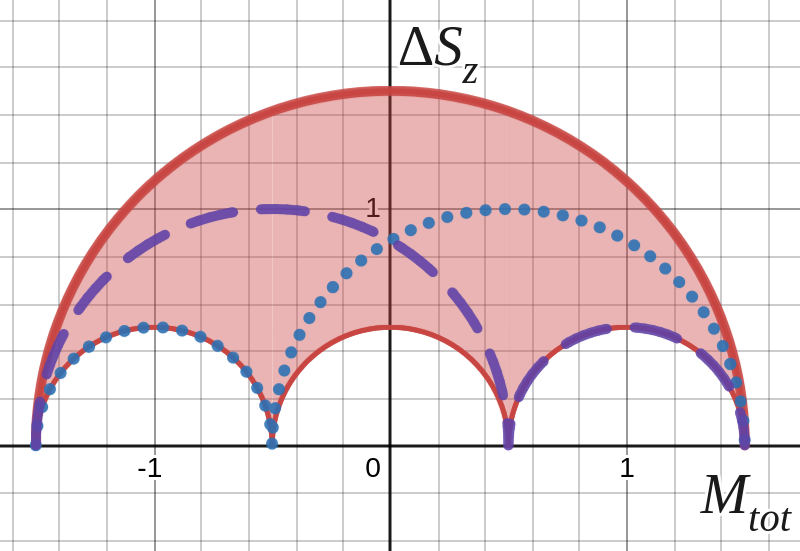}
    \caption{Deviation $\Delta S_z$ versus $M_\tot$, for $S_0=\tfrac{3}{2}$: for the scenario of a minimal $y$-value ($\Delta S_z(y_\m)$, dotted blue) and for the scenario of a maximal $y$-value ($\Delta S_z(y_{\max})$, dashed purple). The minimal deviation, $(\Delta S_z)_\m$ (solid, thin red line) and the maximal deviation $(\Delta S_z)_{\max}$ (solid, thick red line) are brought for comparison.}
    \label{fig:Dsz_min_max_y_min_y_max}
\end{figure}

\begin{figure}
    \centering
    \includegraphics[width=0.45\textwidth]{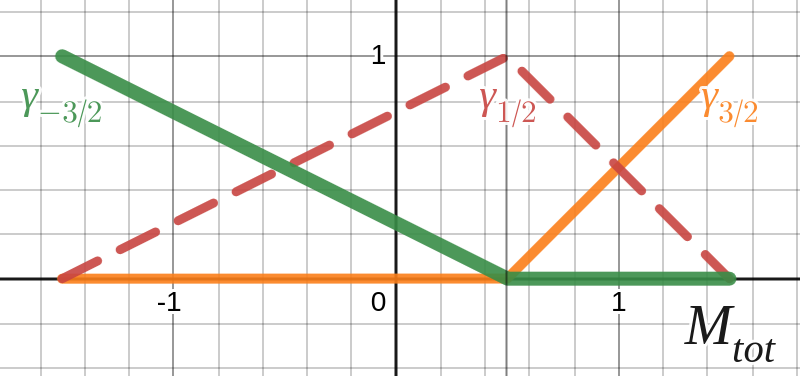}
    \caption{Ensemble coefficients, $\gamma_{-3/2}$ (thick solid green), $\gamma_{1/2}$ (dashed red), and $\gamma_{3/2}$ (solid orange), versus $M_\tot$, for $S_0=\tfrac{3}{2}$, for the scenario of $y_{\max}$. $\gamma_{-1/2}=0$ is not shown. }
    \label{fig:gammas_S_3_2_y_max}
\end{figure}

The lowest value, $y_\m$, can be expressed as
\begin{align}
&y_\m(M_\tot)  = &\lp\{ 
    \begin{array}{lcl}
        -\thalf M_\tot - \frac{3}{4}  & : & M_\tot \in [-\frac{3}{2},-\half]\\[3pt]
        \tfrac{1}{4} M_\tot - \tfrac{3}{8} & : & M_\tot \in [-\half, \tfrac{3}{2}]\\[3pt]
        \end{array}
\right. 
\end{align}
(see Fig.~\ref{fig:y_min_max_0_1}).
Due to the geometry of the $x-y$ domain in this Case, for each value of $y_\m$, there is \emph{one} value of $x$:
\begin{align}
&x_1(M_\tot) \defeq x(y_\m)= \lp\{ 
    \begin{array}{lcl}
        - M_\tot - \thalf  & : & M_\tot \in [-\frac{3}{2},-\half]\\[3pt]
        \thalf M_\tot + \tfrac{1}{4} & : & M_\tot \in [-\half,  \tfrac{3}{2}]\\[3pt]
    \end{array}
\right. 
\end{align}
(see Fig.~\ref{fig:x_min_max_1_2_S_1_5})
As a result, the ensemble coefficients equal
\begin{align}
\gamma_{3/2}   & = \lp\{ 
    \begin{array}{lcl}
        0             & : & M_\tot \in [-\frac{3}{2},-\half]\\[3pt]
        \thalf M_\tot + \tfrac{1}{4}  \mbox{\hspace{1ex}} & : & M_\tot \in [-\half, \frac{3}{2}]
    \end{array}
    \right.  \nonumber \\
\gamma_{1/2}   & = \lp\{ 
    \begin{array}{lcl}
        0 \mbox{\hspace{11ex}} & : & M_\tot \in [-\frac{3}{2},-\half]\\[3pt]
        0  & : & M_\tot \in [-\half, \frac{3}{2}]
    \end{array}
    \right.   \\
\gamma_{-1/2}   & = \lp\{ 
    \begin{array}{lcl}
        \tfrac{3}{2} + M_\tot & : & M_\tot \in [-\frac{3}{2},-\half]\\[3pt]
        - \thalf M_\tot + \tfrac{3}{4}  & : & M_\tot \in [-\half, \frac{3}{2}]
    \end{array}
    \right.  \nonumber \\
\gamma_{-3/2}   & = \lp\{ 
    \begin{array}{lcl}
        -\thalf - M_\tot \mbox{\hspace{2ex}}& : & M_\tot \in [-\frac{3}{2},-\half]\\[3pt]
        0  & : & M_\tot \in [-\half, \frac{3}{2}]
    \end{array}
    \right.  \nonumber 
\end{align}
(see Fig.~\ref{fig:gammas_S_3_2_y_min} and compare to Fig.~\ref{fig:gammas_S_1_5}).
Interestingly, for $M_\tot : -\tfrac{3}{2} \rarr -\thalf$, we observe a linear transition from $\ketbra{\Psi_{N_0,-3/2}}$ to $\ketbra{\Psi_{N_0,-1/2}}$, as in the scenario of $(\Delta S_z)_\m$. But then, for $M_\tot : -\thalf \rarr \tfrac{3}{2}$, we observe a transition from $\ketbra{\Psi_{N_0,-1/2}}$ to $\ketbra{\Psi_{N_0,3/2}}$, skipping $\ketbra{\Psi_{N_0,1/2}}$. This results in $\Delta S_z$ which is higher than the minimal one:
\begin{align}
\Delta S_z (y_\m)   & = \lp\{ 
    \begin{array}{lcl}
        \sqrt{(M_\tot+ \thalf)( \shm M_\tot \shm \tfrac{3}{2})} & : & \!\! M_\tot \!\in\! [\shm \tfrac{3}{2}, \shm \thalf]\\[3pt]
        \sqrt{(M_\tot+ \thalf)(\tfrac{3}{2}-M_\tot)} & : & \!\! M_\tot \!\in\! [\shm \half, \tfrac{3}{2}]
    \end{array}
    \right.
\end{align}
(see Fig.~\ref{fig:Dsz_min_max_y_min_y_max})
Similarly, for the highest possible $y$ value, $y_{\max}$, we obtain the following results:
\begin{align}
&y_{\max}(M_\tot)  = &\lp\{ 
    \begin{array}{lcl}
         \:\:\: \tfrac{1}{4} M_\tot + \tfrac{3}{8}  & : & M_\tot \in [-\frac{3}{2}, \half]\\[3pt]
         -\thalf M_\tot + \frac{3}{4} & : & M_\tot \in [\half, \tfrac{3}{2}]\\[3pt]
    \end{array}
\right. 
\end{align}
\begin{align}
&x_2(M_\tot) \defeq x(y_{\max})= \lp\{ 
    \begin{array}{lcl}
        \tfrac{1}{4} - \thalf M_\tot & : & M_\tot \in [-\frac{3}{2}, \half]\\[3pt]
         -\thalf + M_\tot & : & M_\tot \in [\half, \tfrac{3}{2}]\\[3pt]
    \end{array}
\right. 
\end{align}
(see Figs.~\ref{fig:y_min_max_0_1} and~\ref{fig:x_min_max_1_2_S_1_5}, respectively). 
The ensemble coefficients equal
\begin{align}
\gamma_{3/2}   & = \lp\{ 
    \begin{array}{lcl}
        0                & : & M_\tot \in [-\frac{3}{2},\half]\\[3pt]
        M_\tot - \thalf  & : & M_\tot \in [\half, \frac{3}{2}]
    \end{array}
    \right.  \nonumber \\
\gamma_{1/2}   & = \lp\{ 
    \begin{array}{lcl}
        \tfrac{3}{4} + \thalf M_\tot  & : & M_\tot \in [-\frac{3}{2},\half]\\[3pt]
        \tfrac{3}{2} - M_\tot  & : & M_\tot \in [\half, \frac{3}{2}]
    \end{array}
    \right.   \\
\gamma_{-1/2}   & = \lp\{ 
    \begin{array}{lcl}
        0  \mbox{\hspace{10ex}} & : & M_\tot \in [-\frac{3}{2},\half]\\[3pt]
        0  & : & M_\tot \in [\half, \frac{3}{2}]
    \end{array}
    \right.  \nonumber \\
\gamma_{-3/2}   & = \lp\{ 
    \begin{array}{lcl}
        \tfrac{1}{4} -\thalf M_\tot & : & M_\tot \in [-\frac{3}{2},\half]\\[3pt]
        0  & : & M_\tot \in [\half, \frac{3}{2}]
    \end{array}
    \right.  \nonumber 
\end{align}
(see Fig.~\ref{fig:gammas_S_3_2_y_max} and compare to Fig.~\ref{fig:gammas_S_1_5}).
For $M_\tot : -\tfrac{3}{2} \rarr \thalf$, we observe a linear transition from $\ketbra{\Psi_{N_0,-3/2}}$ to $\ketbra{\Psi_{N_0,1/2}}$ (while skipping $\ketbra{\Psi_{N_0,-1/2}}$), and then, for $M_\tot : \thalf \rarr \tfrac{3}{2}$, from $\ketbra{\Psi_{N_0,1/2}}$ to $\ketbra{\Psi_{N_0,3/2}}$, as in the $(\Delta S_z)_\m$ scenario. This picture is complementary to the one we observed for $y_\m$: this time we skipped $\ketbra{\Psi_{N_0,-1/2}}$. The result for $\Delta S_z$ with maximal $y$ is:
\begin{align}
\Delta S_z (y_{\max})   & = \lp\{ 
    \begin{array}{lcl}
        \sqrt{( \thalf - M_\tot)( M_\tot + \tfrac{3}{2})} & : & \!\! M_\tot \!\in\! [\shm \tfrac{3}{2}, \thalf]\\[3pt]
        \sqrt{(\thalf - M_\tot)( M_\tot - \tfrac{3}{2})} & : & \!\! M_\tot \!\in\! [\half, \tfrac{3}{2}]
    \end{array}
    \right.
\end{align}
(see Fig.~\ref{fig:Dsz_min_max_y_min_y_max}).
Therefore, in Case (d), introduction of a magnetic field and a subsequent minimization of the energy determines both $x$ and $y$, and thus the ground state $\hat{\Gamma}$ unambiguously. 

Finally, we discuss \textbf{Case (e)} of the main text: $N_\tot = N_0 + \alpha$, $\alpha \in [0,1)$, $S_0 = \thalf$ and $S_1 = 1$. This Case corresponds to adding an electron, e.g., to C$^+$ towards C, while
varying $M_\tot$.
In this Case we have five ensemble coefficients, with three constraints (Eq.~(\EqNineMT)). The coefficients can be expressed using two variables, $x$ and~$y$:
\begin{align}
    \gamma_{N_0,1/2}  &= \frac{1-\alpha}{2} + y \nonumber \\
    \gamma_{N_0,-1/2} &= \frac{1-\alpha}{2} - y \nonumber \\
    \gamma_{N_0+1,1}  &= \frac{\alpha-x}{2} + \frac{M_\tot-y}{2} \\ \gamma_{N_0+1,0}  &= x                      \nonumber \\
    \gamma_{N_0+1,-1} &= \frac{\alpha-x}{2} - \frac{M_\tot-y}{2}. \nonumber
\end{align}
With these coefficients, we find that the total energy, $E(N_\tot,M_\tot) = (1-\alpha) E(N_0) + \alpha E(N_0+1)$, and $\stat{\hat{S}_z} = M_\tot$ are determined without ambiguity. In contrast, other quantities of interest depend on $x$ or on $y$.  The even moments of $\hat{S}_z$ equal $\stat{\hat{S}_z^{2k}} = S_0^{2k} + (S_1^{2k} - S_0^{2k}) \alpha - S_1^{2k} x$, and are linear in $x$ and independent of $y$, whereas the odd moments of $\hat{S}_z$ are $\stat{\hat{S}_z^{2k+1}} = S_0^{2k} y + S_1^{2k} (M_\tot-y)$ and are linear in $y$ and independent of $x$. Similarly to Case~(d), to determine the ensemble ground state, we can add two constraints, e.g., for $\stat{\hat{S}_z^2}$ and $\stat{\hat{S}_z^3}$. 

The spin-densities are not uniquely determined. They can be expressed as 
\begin{align}
&n^\s_\ens(\rr) = \half \lp( (1-\alpha) n_{N_0}(\rr) + \alpha n_{N_0+1}(\rr) \rp) + \nonumber \\ 
&+ \half \delta_\s \lp[ M_\tot Q_{N_0+1,1}(\rr) + y \lp( \frac{Q_{N_0,1/2}(\rr)}{1/2} - Q_{N_0+1,1}(\rr)\rp) \rp]
\end{align}
(cf.\ Eq.~\eqref{eq:n_s_ens__d}).
The spin-densities are independent of $x$ and linear in $y$. It directly follows that the total density
\begin{equation}
    n_\ens(N_\tot,M_\tot) = (1-\alpha) n_{N_0}(\rr) + \alpha n_{N_0+1}(\rr),
\end{equation}
is piecewise-linear in $\alpha$ and unambiguously defined, as expected, while the spin distribution is given by
\begin{align}
Q_\ens(\rr) =  (M_\tot-y) Q_{N_0+1,1}(\rr) + y \frac{Q_{N_0,1/2}(\rr)}{1/2}.
\end{align}
It depends only on $y$, and can be determined by setting $\stat{\hat{S}_z^3}$, according to Sec.~\ref{sec:Higher_moments}.

\begin{figure*}
    \centering
    \includegraphics[width=0.24\textwidth]{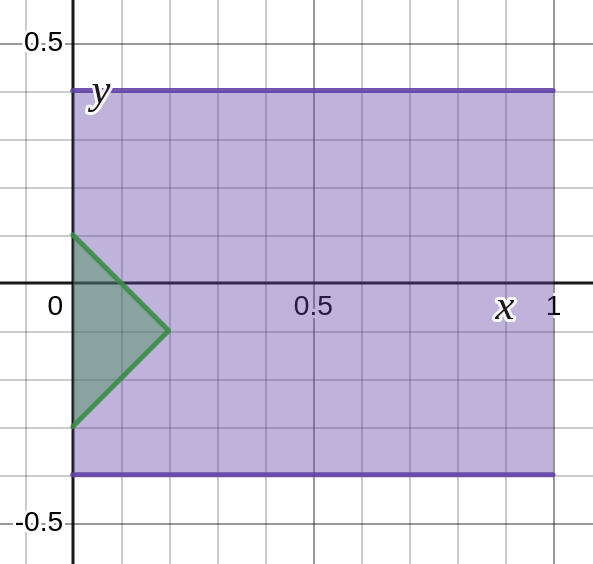}
    \includegraphics[width=0.24\textwidth]{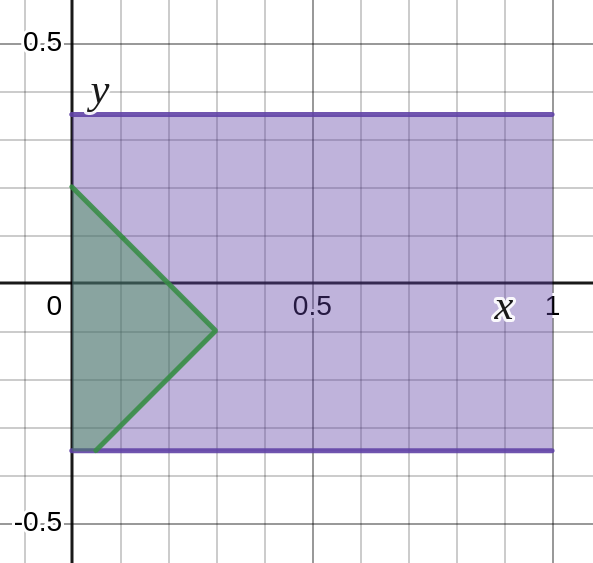}
    \includegraphics[width=0.24\textwidth]{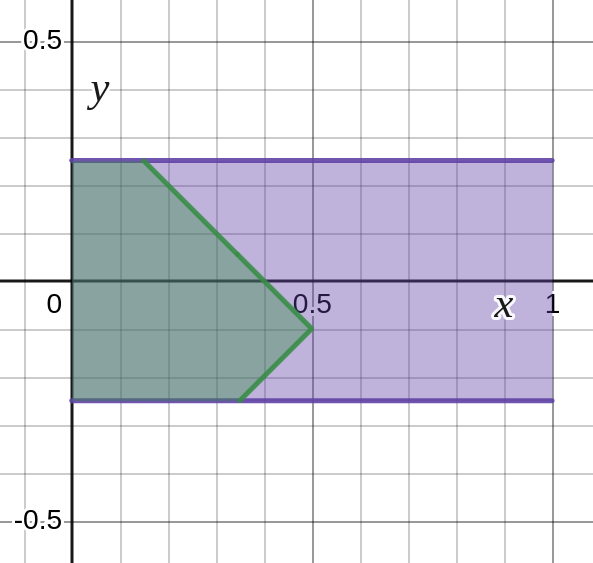}
    \includegraphics[width=0.24\textwidth]{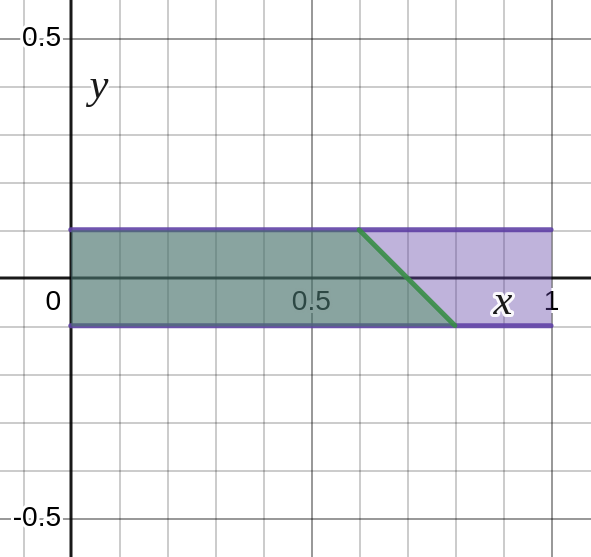}
    \caption{The allowed domain of $x$ and $y$, for Case (e) ($N_\tot = N_0+\alpha$, $S_0=\tfrac{1}{2}$, $S_1=1$), is the intersection of the green and purple areas. In all panels, $M_\tot=-0.1$, while $\alpha = 0.2$, 0.3, 0.5 and 0.8 (left to right). For an interactive plot, see: \url{https://www.desmos.com/calculator/rtvugkfxap}. }
    \label{fig:Case_e_x_y_domain}
\end{figure*}

As in previous Cases, we address the scenario of minimal $\Delta S_z$. Here, $\Delta S_z = \sqrt{\tfrac{1}{4}(1+3\alpha) -x -M_\tot^2}$; a minimum is obtained for the highest possible $x$. The domain of $x$ and $y$, which emerges from the requirement that all the ensemble coefficients are within $[0,1]$, can be described by the inequalities: $ 0 \leqs x \leqs 1$, $\abs{y}\leqs \thalf(1-\alpha)$ and $|y-M_\tot| \leqs \alpha - x$. The $x-y$ domain is illustrated in Fig.~\ref{fig:Case_e_x_y_domain}. 

\begin{figure}
    \centering
    \includegraphics[width=0.37\textwidth]{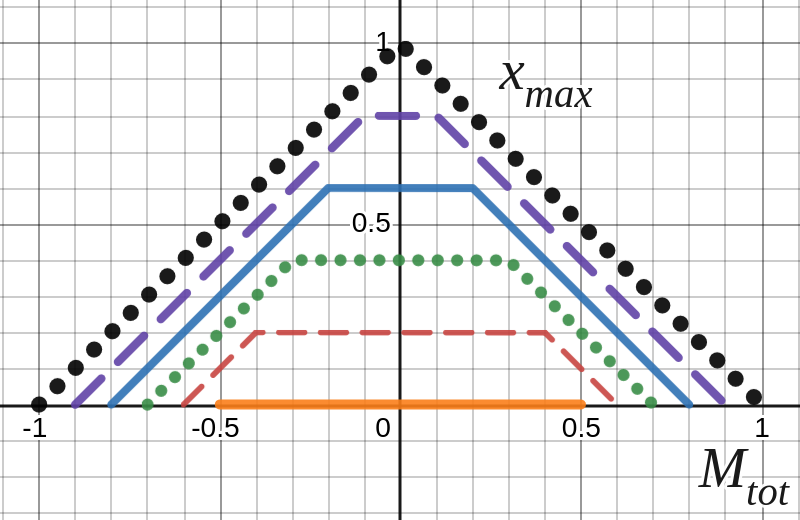}
    \caption{Dependence of the maximal value of $x$ ($x_{\max}$) on $M_\tot$, for different values of $\alpha$: $\alpha=0$ -- solid orange, $\alpha=0.2$ -- dashed red, $\alpha=0.4$ -- dotted green, $\alpha=0.6$ -- solid blue, $\alpha = 0.8$ -- dashed purple, $\alpha=1$ -- dotted black.}
    \label{fig:Case_e_x_max}
\end{figure}

\begin{figure}
    \centering
    \includegraphics[width=0.40\textwidth]{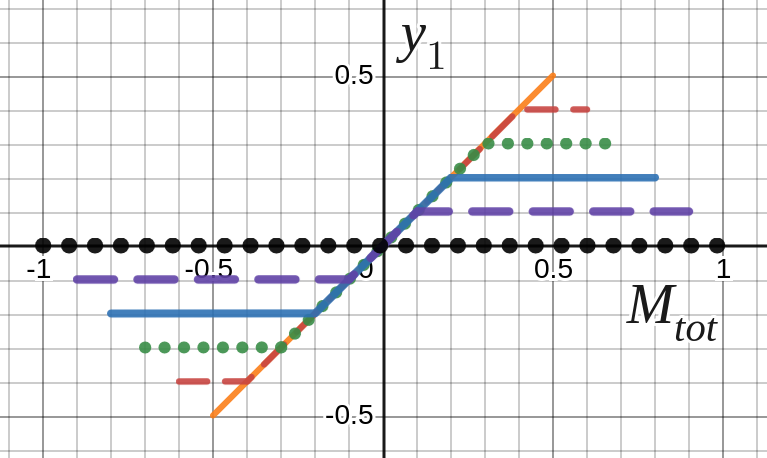}
    \caption{Dependence of $y_1$ -- the $y$-value that corresponds to the maximal value of $x$, on $M_\tot$, for different values of $\alpha$. Color coding as in Fig.~\ref{fig:Case_e_x_max}.}
    \label{fig:Case_e_y_1}
\end{figure}

In Case (e), the highest value of $x$ is
\begin{align}
    x_{\max}(M_\tot) = \lp\{ 
        \begin{array}{rcl}
                \tfrac{1+\alpha}{2} + M_\tot  & : & M_\tot \in [-\tfrac{1+\alpha}{2}, -\tfrac{1-\alpha}{2}]\\[3pt]
                \alpha               & : & M_\tot \in [-\tfrac{1-\alpha}{2}, \tfrac{1-\alpha}{2}]\\[3pt]
                \tfrac{1+\alpha}{2} - M_\tot  & : & M_\tot \in [\tfrac{1-\alpha}{2}, \tfrac{1+\alpha}{2}]
              \end{array}.
           \right.  
\end{align}
As in Case (d), also here for each value of $x_{\max}$ there is \emph{one} corresponding value of $y$:
\begin{align}
    y_1(M_\tot) \defeq y(x_{\max}) = \lp\{ 
\begin{array}{rcl}
                - \tfrac{1 - \alpha}{2} & : & M_\tot \in [-\tfrac{1+\alpha}{2}, -\tfrac{1-\alpha}{2}]\\[3pt]
                M_\tot               & : & M_\tot \in [-\tfrac{1-\alpha}{2}, \tfrac{1-\alpha}{2}]\\[3pt]
                \tfrac{1 - \alpha}{2}  & : & M_\tot \in [\tfrac{1-\alpha}{2}, \tfrac{1+\alpha}{2}]
              \end{array}
           \right.  
\end{align}
The dependence of $x_{\max}$ and $y_1$ on $M_\tot$, for different values of $\alpha$, is depicted on Figs.~\ref{fig:Case_e_x_max} and \ref{fig:Case_e_y_1}, respectively.
We note in passing that, in contrast to Case (d), \emph{maximizing} $\Delta S_z$ does not uniquely determine the ground state. The lowest possible $x$, $x_\m = 0$, is always available, and corresponds to a range of $y$ values.

\begin{figure*}
    \centering
    \includegraphics[width=0.19\textwidth]{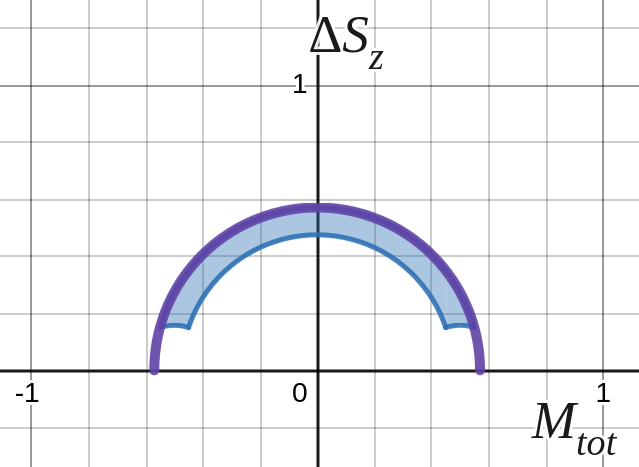}
    \includegraphics[width=0.19\textwidth]{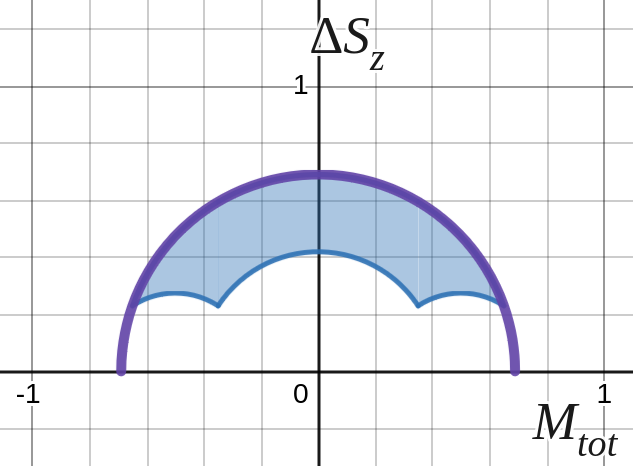}
    \includegraphics[width=0.19\textwidth]{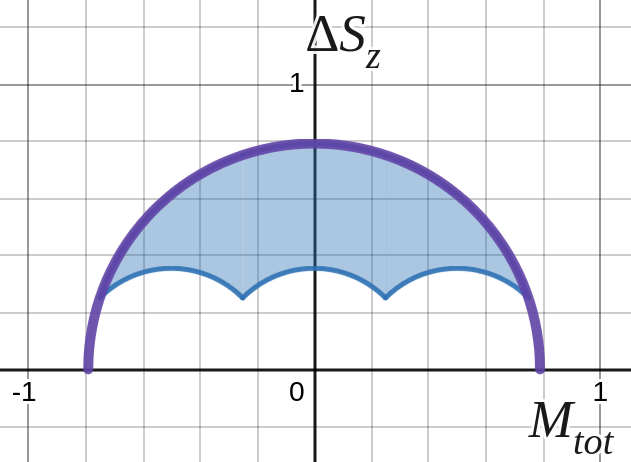}
    \includegraphics[width=0.19\textwidth]{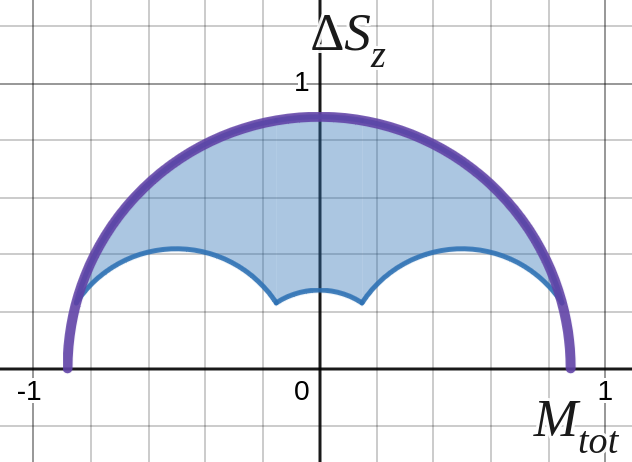}
    \includegraphics[width=0.19\textwidth]{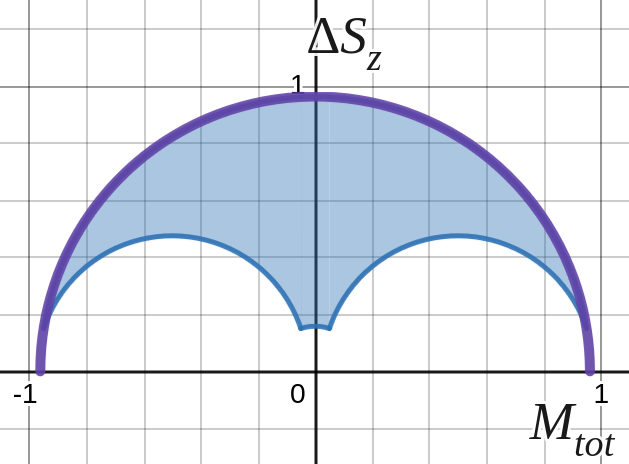}
    \caption{Deviation $\Delta S_z$ versus $M_\tot$, for Case (e), for different values of $\alpha$. From left to right: $\alpha = 0.1$, 0.3, 0.5, 0.7, 0.9. Minimal deviation, $(\Delta S_z)_\m$: solid blue line, maximal deviation,  $(\Delta S_z)_{\max}$: thick solid purple line. For an interactive plot, see: \url{https://www.desmos.com/calculator/msmk9qqtuu}.}
    \label{fig:Case_e_DSz_min}
\end{figure*}

The minimal value of $\Delta S_z$ is 
\begin{align}
    &(\Delta S_z)_\m = \nonumber \\
    &\lp\{ 
        \begin{array}{lcl}
                \sqrt{\frac{\alpha-1}{4}-M_\tot(1+M_\tot)}   & : & M_\tot  \in  [- \frac{1+\alpha}{2}, - \frac{1-\alpha}{2}]\\
                \sqrt{\tfrac{1-\alpha}{4} - M_\tot^2}  & : & M_\tot \in  [-\frac{1-\alpha}{2}, \frac{1-\alpha}{2}]\\
                \sqrt{M_\tot(1-M_\tot)-\frac{1-\alpha}{4}}   & : & M_\tot \in   [\frac{1-\alpha}{2}, \frac{1+\alpha}{2}]
              \end{array}
           \right.,  \nonumber \\
\end{align}
whereas the maximal value is $(\Delta S_z)_{\max{}} = \sqrt{\tfrac{1}{4}(1+3\alpha) - M_\tot^2}$; see Fig.~\ref{fig:Case_e_DSz_min}. The particular cases for $\alpha=0$ (Case (a)) and $\alpha=1$ (Case (c)) are in agreement with derivations above.

For the scenario of minimal $\Delta S_z$, the ensemble coefficients equal
\begin{align}
\gamma_{N_0,1/2}   & = \lp\{ 
    \begin{array}{lcl}
        0             & : & M_\tot \in [-\frac{1+\alpha}{2},-\frac{1-\alpha}{2}]\\[3pt]
        \tfrac{1-\alpha}{2} + M_\tot & : & M_\tot \in [-\frac{1-\alpha}{2},\frac{1-\alpha}{2}]\\[3pt]
        1-\alpha  & : & M_\tot \in [\frac{1-\alpha}{2},\frac{1+\alpha}{2}]
    \end{array}
                \right.  \nonumber \\
\gamma_{N_0,-1/2}   & = \lp\{ 
    \begin{array}{lcl}
        1-\alpha  & : & M_\tot \in [-\frac{1+\alpha}{2},-\frac{1-\alpha}{2}]\\[3pt]
        \tfrac{1-\alpha}{2} - M_\tot  & : & M_\tot \in [-\frac{1-\alpha}{2},\frac{1-\alpha}{2}]\\[3pt]
        0  & : & M_\tot \in [\frac{1-\alpha}{2},\frac{1+\alpha}{2}]
    \end{array}
                \right.  \nonumber \\
\gamma_{N_0+1,1}   & = \lp\{ 
    \begin{array}{lcl}
        0             & : & M_\tot \in [-\frac{1+\alpha}{2},-\frac{1-\alpha}{2}]\\[3pt]
        0             & : & M_\tot \in [-\frac{1-\alpha}{2},\frac{1-\alpha}{2}]\\[3pt]
        M_\tot - \frac{1-\alpha}{2}  & : & M_\tot \in [\frac{1-\alpha}{2},\frac{1+\alpha}{2}]
    \end{array}
                \right.  \nonumber \\
\gamma_{N_0+1,0}   & = \lp\{ 
    \begin{array}{lcl}
        \tfrac{1+\alpha}{2}+M_\tot    & : & M_\tot \in [-\frac{1+\alpha}{2},-\frac{1-\alpha}{2}]\\[3pt]
        \alpha             & : & M_\tot \in [-\frac{1-\alpha}{2},\frac{1-\alpha}{2}]\\[3pt]
        \tfrac{1+\alpha}{2}-M_\tot  & : & M_\tot \in [\frac{1-\alpha}{2},\frac{1+\alpha}{2}]
    \end{array}
                \right.  \nonumber \\
\gamma_{N_0+1,-1}   & = \lp\{ 
    \begin{array}{lcl}
       - M_\tot - \frac{1-\alpha}{2}            & : & M_\tot \in [-\frac{1+\alpha}{2},-\frac{1-\alpha}{2}]\\[3pt]
        0             & : & M_\tot \in [-\frac{1-\alpha}{2},\frac{1-\alpha}{2}]\\[3pt]
        0  & : & M_\tot \in [\frac{1-\alpha}{2},\frac{1+\alpha}{2}]
    \end{array}.
                \right.  \nonumber \\
\end{align}
For any value of $M_\tot$ we have three nonzero coefficients, at most. 

Finally, the spin distribution for the case of minimal $\Delta S_z$ equals
\begin{align}
&Q_\ens(\rr)  = \nonumber\\
&\!\!\lp\{ 
    \begin{array}{lcl}
       \!\!\! (M_\tot \shp \frac{1 \shm \alpha}{2}) Q_{N_0+1,1}(\rr) \!\shm\! \frac{1 \shm \alpha}{2}  \frac{Q_{N_0,1/2}(\rr)}{1/2} \!\!\!\!& : &\!\!\! M_\tot \!\!\in\!\! [\shm \frac{1+\alpha}{2}, \shm \frac{1 \shm \alpha}{2}]\\[3pt]
        \!\!\! M_\tot \frac{Q_{N_0,1/2}(\rr)}{1/2} \!\!& : &\!\!\! M_\tot \!\in\! [\shm \frac{1 \shm \alpha}{2},\frac{1 \shm \alpha}{2}]\\[3pt]
        \!\!\! (M_\tot \!\shm\! \frac{1 \shm \alpha}{2}) Q_{N_0+1,1}(\rr) \shp \frac{1 \shm \alpha}{2}  \frac{Q_{N_0,1/2}(\rr)}{1/2} \!\!\!\! & : & \!\! M_\tot \!\in\! [\frac{1 \shm \alpha}{2},\frac{1+\alpha}{2}]
    \end{array}.
\right. 
\end{align}
Interestingly, also here, $Q_\ens(\rr)$, and therefore $\zeta_\ens(\rr)$, are piecewise-linear in both $M_\tot$ and in $\alpha$.

In presence of a magnetic field $\mathbf{B}(\rr) = B_0 f(\rr) \hat{z}$ in Case (e), the total energy becomes $\tilde{E}_\ens(N_\tot,M_\tot) = (1-\alpha) E(N_0) + \alpha E(N_0+1) +  \muB B_0 M_\tot F_{N_0+1,1} +  \muB B_0 (F_{N_0,1/2}-F_{N_0+1,1}) y$. Also here we are looking then for the lowest and the highest possible values of $y$. 

Analyzing the $x-y$ domain of Case (e), we found that the lowest value, $y_\m$, can be expressed as
\begin{align}
&y_\m(M_\tot)  = &\lp\{ 
    \begin{array}{lcl}
        -\tfrac{1-\alpha}{2} & : & M_\tot \in [-\frac{1+\alpha}{2}, \tfrac{3\alpha-1}{2}]\\[3pt]
        M_\tot - \alpha  & : & M_\tot \in [\frac{3\alpha-1}{2},\frac{1+\alpha}{2}]\\[3pt]
        \end{array}
\right. .
\end{align}
\begin{figure}
    \centering
    \includegraphics[width=0.45\textwidth]{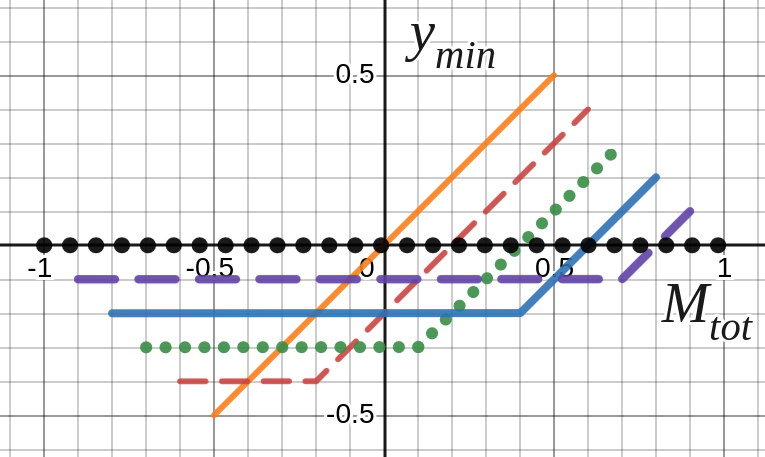}
    \caption{Dependence of $y_\m$, the minimal value
of $y$ in Case (e), on $M_\tot$, for different values of $\alpha$. Color coding as in Fig.~\ref{fig:Case_e_x_max}.}
    \label{fig:Case_e_y_min}
\end{figure}
\begin{figure}
    \centering
    \includegraphics[width=0.45\textwidth]{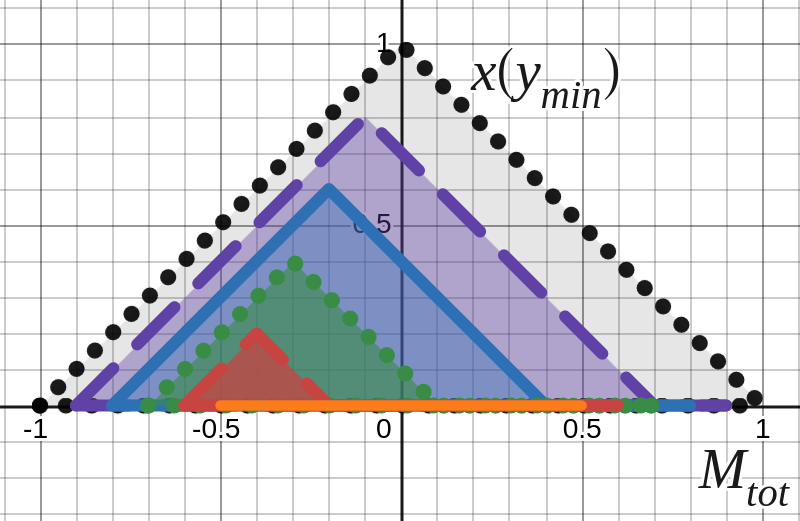}
    \caption{Dependence of the allowed range of $x$-values, which correspond to the minimal $y$-value, in Case (e), on $M_\tot$. Color coding for different values of $\alpha$ as in Fig.~\ref{fig:Case_e_x_max}. The range of allowed $x$-values is reflected by the shaded area on the plot.}
    \label{fig:Case_e_x_of_y_min}
\end{figure}
Importantly, in this Case, $x$ is not determined uniquely: for $M_\tot \in [-\frac{1+\alpha}{2}, -\tfrac{1-\alpha}{2}]$, we find that $x \in [0, M_\tot+\frac{1+\alpha}{2}]$,  for $M_\tot \in [-\frac{1-\alpha}{2}, \tfrac{3\alpha-1}{2}]$, we find that $x \in [0, -M_\tot+\frac{3\alpha-1}{2}]$, and for $M_\tot \in [ \tfrac{3\alpha-1}{2}, \frac{1+\alpha}{2},]$, $x=0$. Figs.~\ref{fig:Case_e_y_min} and~\ref{fig:Case_e_x_of_y_min} show $y_\m$ and the range of $x$ graphically.
Therefore, while the magnetic field in this case can set the spin-densities, the odd spin moments, and other quantities, the ground state remains ambiguous (due to~$x$).

\bibliography{bib_EK_2022}